\documentclass[journal,twoside,web]{ieeecolor}

\usepackage{generic}
\usepackage{cite}
\usepackage{textcomp}

\usepackage{times}

\usepackage{caption}

\usepackage[siunitx]{circuitikz}

\usepackage{microtype}

\usepackage[hidelinks]{hyperref}

\usepackage{amsfonts}

\usepackage{amsthm}
\usepackage{amssymb}
\usepackage{amsmath}
\usepackage{extarrows}
\usepackage{enumerate}
\usepackage{graphicx}
\usepackage{subfigure}
\usepackage{pgf,pgfarrows,pgfnodes,pgfautomata,pgfheaps}
\usepackage{tikz}
\usetikzlibrary{arrows,automata,fit,matrix}
\usepackage{algorithm,algorithmic}
\usepackage{booktabs}
\usepackage{verbatim}
\usepackage{bm}
\usepackage{bbm}
\usepackage{epstopdf}

\usepackage{mathtools}

\usepackage{url}

\usepackage[off,crop=off]{auto-pst-pdf}

\usepackage{mhchem}

\DeclareMathOperator*{\argmin}{arg\,min}

\newcommand{\RE}{\mathbb{R}}
\newcommand{\NA}{\mathbb{N}_0}

\newcommand{\calE}{\mathcal{E}}
\newcommand{\calG}{\mathcal{G}}
\newcommand{\calH}{\mathcal{H}}

\newcommand{\calR}{\mathcal{R}}
\newcommand{\calS}{\mathcal{S}}

\newcommand{\lb}{\underline{\alpha}}
\newcommand{\ub}{\overline{\alpha}}

\newcommand{\dsl}{\displaystyle}

\newcommand{\pa}{\mathcal{H}}
\newcommand{\upp}[1]{{{#1}^{\uparrow}}}

\newcommand{\frR}{\mathfrak{R}}        
\newcommand{\rr}{\mathbf{rr}}        

\newcommand{\act}[1]{\xlongrightarrow{#1}}          

\newtheorem{definition}{Definition}

\newcommand{\ha}{\hat{\alpha}}

\newcommand{\hf}{\hat{f}}

\newcommand{\hp}{\hat{p}}

\newcommand{\hq}{\hat{q}}
\newcommand{\hJ}{\hat{J}}
\newcommand{\hK}{\hat{K}}
\newcommand{\hL}{\hat{L}}
\newcommand{\hR}{\hat{\mathcal{R}}}
\newcommand{\hS}{\hat{\mathcal{S}}}
\newcommand{\hv}{\hat{v}}

\newcommand{\hX}{\hat{X}}

\newcommand\deq{\stackrel{\mathclap{\tiny\mbox{d}}}{=}}

\newcommand{\rI}{%
  \textup{\uppercase\expandafter{\romannumeral 1}}%
}

\newcommand{\rII}{%
  \textup{\uppercase\expandafter{\romannumeral 2}}%
}

\newtheorem{example}{Example}

\newtheorem{proposition}{Proposition}
\newtheorem{theorem}{Theorem}
\newtheorem{remark}{Remark}

\usepackage{array}
\newcolumntype{H}{>{\setbox0=\hbox\bgroup}c<{\egroup}@{}}

\title{\LARGE \bf
Optimality-preserving Reduction of Chemical Reaction Networks
}

\author{Kim G. Larsen$^{1}$, Daniele Toller$^{1}$, Mirco Tribastone$^{2}$, Max Tschaikowski$^{1}$ and Andrea Vandin$^{3}$
\thanks{$^{1}$ K. G. Larsen, D. Toller and M. Tschaikowski are with Aalborg University, Denmark}
\thanks{$^{2}$ M. Tribastone is with IMT Lucca, Italy}
\thanks{$^{3}$ A. Vandin is with Sant'Anna Pisa, Italy and DTU Compute, Denmark}
}

\usepackage{numprint}

\begin{document}

\maketitle
\thispagestyle{empty}
\pagestyle{empty}

\begin{abstract}
Across many disciplines, chemical reaction networks (CRNs) are an established population model defined as a system of coupled nonlinear ordinary differential equations. In many applications, for example, in systems biology and epidemiology,  CRN parameters such as the kinetic reaction rates can be used as control inputs to steer the system toward a given target. Unfortunately, the resulting optimal control problem is nonlinear, therefore, computationally very challenging. We address this issue by introducing an optimality-preserving reduction algorithm for CRNs. The algorithm partitions the original state variables into a reduced set of macro-variables for which one can define a reduced optimal control problem from which one can exactly recover the solution of the original control problem. Notably, the reduction algorithm runs with polynomial time complexity in the size of the CRN.  We use this result to reduce reachability and control problems of large-scale protein-interaction networks and vaccination models with hundreds of thousands of state variables.
\end{abstract}

\section{Introduction}\label{sec:intro}


The interplay between control theory and systems biology is instrumental to gain insights into the dynamics of natural systems across different scales (e.g.,~\cite{doi:10.1073/pnas.0308265100}). In particular, the problem of controlling a biological system is relevant in applications such as smart therapeutics and biosensors~\cite{NatureNanoTechDNA}. Mathematically, this can be studied as the problem of controlling a formal chemical reaction network (CRN), whereby the biological system under study is modeled as a (finite) set of species that interact across a (finite) set of reaction channels. This representation admits both a stochastic interpretation in terms of a continuous-time Markov chain (CTMC), where discrete changes in the population levels of each species are tracked, and a deterministic one as a system of nonlinear ordinary differential equations (ODEs), where each equation tracks the time evolution of the concentration of each species. Notably, and particularly relevant for the theoretical developments in this paper, under mild conditions the deterministic equations correspond to a limit regime of a family of CTMCs (e.g.,~\cite{kurtz-chem}).

In this setting, the control inputs may be represented by the parameter values of designated reaction rates~\cite{NatureBioControl}, such that the overall controller design may be studied as an optimal control problem. Treating certain rates as inputs can also be used for the complementary goal of studying the open-loop behavior of the system when some parameters are unknown/uncertain, by estimating reachable sets~\cite{DBLP:journals/automatica/Lygeros04}; this is a pressing problem in systems biology, where rate parameters are often not directly accessible.

Controlling the biological system by studying its ODEs in place of the CTMC is appealing because the ODE system size has, in general, exponentially fewer equations.
%
%
%
%
However, the control problem is computationally prohibitive in general due to the fact that it is nonlinear~\cite{DBLP:journals/tac/PappasLS00,DBLP:conf/rtss/ChenAS12}.
One approach to tackling this problem is to devise  an \emph{optimality-preserving} reduction of a control system, where the hope is to solve a reduced optimal control problem instead of the original one. While for linear systems this problem is well-understood~\cite{DBLP:journals/tac/PappasLS00}, it remains challenging for nonlinear ones. 

In this paper we consider CRNs with the well-known mass-action semantics (e.g.,~\cite{10.1371/journal.pcbi.1004012,NACO2019}), leading to ODE systems with polynomial right-hand sides. Here, reactions are characterized by rate parameters which can be used as inputs, taken from bounded domains. The optimal control problem consists in finding the values of those parameters such that a given cost function is minimized. We present an optimality-preserving reduction method based on a partition of the set of species, thus corresponding to a partition of the set of ODE variables.  This follows a long tradition in the development of \emph{lumping} techniques for (bio-)chemical systems (e.g.,~\cite{okino1998,Snowden:2017aa}), most of which are concerned with preserving the dynamics of the system and not of the solution of the control problem as done here.

\begin{figure}[tp]
    \begin{center}
        \begin{tikzpicture}[<->,>=stealth',shorten >=1pt,auto,semithick]
        \matrix[row sep=1cm, column sep=1cm]
        {
        \pgfmatrixnextcell \node (1) {$\substack{\text{\normalsize CTMC family} \\ \text{\normalsize of CCRN}}$};
        \pgfmatrixnextcell \node (2) {$\substack{\text{\normalsize CTMC family} \\ \text{\normalsize of lumped CCRN}}$}; \\
        \pgfmatrixnextcell \node (3) {$\substack{\text{\normalsize Control system} \\ \text{\normalsize of CCRN}}$};
        \pgfmatrixnextcell \node (4) {$\substack{\text{\normalsize Control system} \\ \text{\normalsize of lumped CCRN}}$}; \\
        };
        \path (1) edge node {(2)} (2)
            (3) edge node [dashed] {(3)} (4)
            (3) edge node {Fluid limit (1)} (1)
            (2) edge node {Fluid limit (1)} (4);
        \end{tikzpicture}
    \end{center}
\caption{Visualization of the main result. As first, we approximate the control systems of the original and the lumped CCRN by means of suitable CTMC families (1). Then, we show that the lumped CTMC family admits the same optimal value as the original one (2). Combining (1) and (2), we conclude that the lumping of the control system is optimality preserving (3).}\label{fig:proof:intro}
\end{figure}
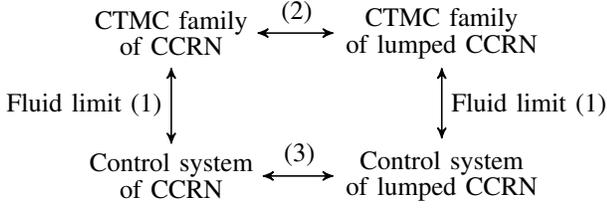

Our reduction is exact  in the sense that one can define a reduced optimal control problem whose solution can be exactly related to that of the original problem. Based on this, we develop an algorithm that finds the coarsest partition, i.e., the maximal lumping, that satisfies this property. The algorithm is based on previous recent results for lumping of uncertain Markov chains (essentially seen as linear control systems~\cite{antoulas2005approximation}). Similarly to that, the number of required computational steps is at most polynomial in the number of species and reactions of the CRN. However, the technical machinery required here is profoundly different and, importantly, identifies the polynomial ODE system of a mass-action CRN as the deterministic limit process of a family of CTMCs. Specifically, in the derivation of the main result, visualized in Fig.~\ref{fig:proof:intro}, we:
\begin{itemize}
    \item make use of fluid limit results~\cite{kurtz-chem,dsn16BortolussiGast} and associate to each CCRN a family of continuous-time Markov chains (CTMCs) which, roughly speaking, converge in probability to the control system of the CCRN; 
    \item show that the original CTMC family can be replaced by the CTMC family of a lumped CCRN while preserving optimality; 
    \item show that the control system of the original and the lumped CCRN have common optimal values;
    \item show how an optimal control of the original CCRN can be computed from an optimal control of the lumped CCRN.
\end{itemize}
By doing so, we circumvent the problem of having to relate nonlinear control systems directly.


We implement the aforementioned lumping algorithm in the software tool ERODE~\cite{tacas2017} and apply the theory to two families of case studies. In the first class, we study epidemiological models over weighted networks~\cite{pastor2015epidemic}, where a) each node is subject to vaccination control or b) the network weights are subject to uncertainty. In the second class, we study protein-interaction networks where proteins can bind to the binding sites of a substrate. Here, we show that lumping algorithms developed for autonomous ODE systems (e.g.~\cite{Feret21042009,PNAScttv,poplTCS}) carry over to the case where kinetic parameters are assumed to belong to common intervals, rather than have a common value. Both classes show how one can reduce nonlinear optimal control~\cite{DBLP:journals/automatica/Lygeros04,Liberzon,DBLP:journals/tac/WhitbyCKLTT22} and verification problems~\cite{DBLP:conf/rtss/ChenAS12,Althoff2015a,DBLP:conf/qest/CardelliTTV18} with thousands of state variables on common hardware.


\emph{Related work.} While results on exact optimality-preserving lumping techniques for linear control systems have been explored (see~\cite{DBLP:journals/tac/PappasLS00,antoulas2005approximation} and references therein), nonlinear counterparts are scarce. Bisimulation/abstraction~\cite{bisimulation_lin_sys_Schaft,DBLP:journals/tac/PappasS02,DBLP:journals/automatica/TabuadaP03} is closely related but complementary to CCRN species equivalence. Specifically, for a given observation, the largest bisimulation gives rise to a lumped dynamical system which coincides with the original one up to a previously chosen observation map. Instead, CCRN species equivalence seeks to find from a family of linear observation maps the one that gives rise to the largest bisimulation. 
Since only observation maps expressible by equivalence relations are considered, the coarsest CCRN species equivalence can be computed in polynomial time. Apart from bisimulation/abstraction, we mention decoupling~\cite{DBLP:conf/cdc/ChenT15} that yields substantial speed-ups but may impose restrictive symmetry constraints. While this can be addressed by decoupling approaches~\cite{DBLP:conf/cdc/ChenHT16}, the corresponding lumping is approximative. It is worth noting that our approach is reminiscent to Koopman operator theory which expresses a nonlinear system via an infinite linear one~\cite{rowley2009spectral}. Fluid limits are however complementary to Koopman operator theory. This is because they rely upon probabilistic arguments and their linear (Markov chain) approximations hold true for arbitrary initial conditions, rather than specific ones~\cite{rowley2009spectral}.


\emph{Paper outline.} After reviewing CRNs, Section~\ref{sec:ccrn} introduces controlled CRNs (CCRNs). Section~\ref{sec:ucrn:lump} instead reviews CRN species equivalence from~\cite{DBLP:journals/bioinformatics/CardelliPTTVW21} and extends it to CCRNs. Building upon Section~\ref{sec:ucrn:lump}, Section~\ref{sec:ucrn:lump:fluid} establishes that CCRN species equivalence allows for optimality-preserving lumping of fluid models, while Section~\ref{sec:eval} presents applications in nonlinear system verification and control. The paper concludes in Section~\ref{sec:conc}, while Section~\ref{sec:postponed} contains a proof that is postponed for the benefit of presentation.

\begin{figure}
\begin{center}
\begin{tabular}{r p{6cm}}
$(\calS,\calR_\alpha)$ & CRN with species $\calS$ and reactions $\calR_\alpha$; any reaction $r \in \calR_{\alpha}$ is annotated by reaction coefficient $\alpha_{i_r}$, while $\alpha = (\alpha_{i_r})_{r \in \calR}$ \\
$r = \pi \xrightarrow{\alpha_{i_r}} \rho$ & reaction where multisets $\rho, \pi \in \mathbb{N}_0^\calS$ denote the reactant and the product  \\
$\alpha_i$ & abbreviation for $\alpha_{i_r}$ \\
$\lb,\ub$ & bounds $\lb, \ub \in \RE_{\geq0}^\calR$ satisfying $\lb \leq \ub$ \\
$(\calS,\calR_{[\lb;\ub]})$ & CCRN: as CRN but reactions $r \in \calR_{[\lb;\ub]}$ are annotated by intervals $[\lb_{i_r};\ub_{i_r}]$ \\
$q(\sigma,\theta)$ & transition rate from CTMC state $\rho$ into state CTMC $\theta$, where $\rho, \theta \in \mathbb{N}_0^\calS$ \\
$\big(\NA^\calS,[q_{\lb} ; q_{\ub}]\big)$ & stochastic control system of a CCRN: a family of CTMCs whose time-varying transition rates satisfy $q(\cdot) \in [q_{\lb} ; q_{\ub}]$ \\
$p$ & probability distribution over $\NA^\calS$ \\
$q$ & measurable function $q(\cdot) \in [q_{\lb} ; q_{\ub}]$ unless stated otherwise \\
$X^q$ & element of $\big(\NA^\calS,[q_{\lb} ; q_{\ub}]\big)$ \\
$X_N$ & $N$-th approximation of a CRN or CCRN \\
$\partial_t v^\alpha = f(v^\alpha,\alpha)$ & deterministic control system of a CCRN: a family of ODEs parameterized by time-varying reaction coefficients  $\alpha(\cdot) \in [\lb ; \ub]$ \\
$\alpha$ & measurable function $\alpha(\cdot) \in [\lb ; \ub]$ unless otherwise stated or part of $\calR_\alpha$ \\
$\calE$ & (species) equivalence relation over $\calS$ \\
$\calH$ & partition of $\calS$, usually $\calH = \calS/\calE$ \\
$\upp{\calE}$ & multiset lifting of $\calE$ to $\NA^\calS$ \\
$\upp{\calH}$ & quotient $\NA^\calS / \upp{\calE}$ \\
$H, \upp{H}$ & element of $\calH$ and $\upp{\calH}$, respectively \\
hat $\widehat{\cdot}$ & lumped CCRN, e.g., $\hS$, $\hR$, $\hat{\alpha}$, $\hq$, $\hX$, etc.
\end{tabular}
\caption{A table of frequently used symbols.}
\end{center}
\end{figure}

\section{Controlled CRNs}\label{sec:ccrn}

A mass-action CRN is $(\calS, \calR_\alpha)$ where $\calS$ is a set of species, $\calR$ is a set of reactions and $\alpha = (\alpha_{i_r})_{r \in \calR}$ is a set of kinetic parameters, with $\alpha_{i_r} \geq 0$. Each reaction $r = \pi \xrightarrow{\alpha_{i_r}} \rho$ comprises multisets $\pi$ and $\rho$ of species, denoting the reactants and products, respectively. Mass-action CRNs are traditionally given both a stochastic and a deterministic interpretation as a Markov jump process and a system of polynomial differential equations.



In the stochastic interpretation~\cite{kurtz-chem}, a state of the underlying Markov chain is a species multiset $\sigma \in \mathbb{N}_0^\calS$ giving the number of molecules $\sigma(A)$ for each species $A \in \calS$.  The forward equations are given by the initial value problem 
\begin{equation}\label{eq:forward}
\partial_t p_\sigma = \sum_{\theta} q(\sigma, \theta) p_\sigma ,
\end{equation}
where $p(0)$ is the initial probability measure, whereas the transition rate from state $\sigma$ to state $\theta$ is 
\begin{align}\label{eq:q:const}
q(\sigma, \theta) = q_{\alpha}(\sigma, \theta) := \sum_{ r = \substack{\rho \act{\alpha_{i_r}} \pi \in \calR_\alpha \\ \theta = \sigma + \pi - \rho } } \alpha_{r_i} \cdot \binom{\sigma}{\rho}
\end{align}
We denote by $q = \left(q(\sigma, \theta)\right)_{\sigma,\theta}$ the \emph{transition rate matrix}, where $q(\sigma,\sigma) = -\sum_{\theta \neq \sigma} q(\sigma,\theta)$. The dynamics can be described as follows: when in state $\sigma$, every reaction determines a possible jump that consumes molecules according to the multiplicities of the reactants and yields new molecules according to the products; the reaction fires proportionally (via the kinetic rate $\alpha_i$) to the total number of possible encounters between single molecules of the reacting species. With this in place, the CTMC described by~(\ref{eq:forward}) is denoted by $(X^{q}(t))_{t \geq 0}$.

In the deterministic interpretation~\cite{10.2307/3212147}, instead, the model is described by the system of polynomial differential equations $\partial_t v = f(v,\alpha)$, where the vector field $f : \RE_{\geq0}^\calS \times \RE_{\geq0}^{|\calR|} \to \RE^\calS$ is given, for any species $A \in \calS$, by
\begin{align}\label{eq:det}
f_A(v,\alpha) \! := \!\! \sum_{r = \rho \act{\alpha_{r_i}} \pi \in \calR_\alpha} \alpha_{r_i} (\pi(A) - \rho(A)) \prod_{B \in \calS} \frac{v_B^{\rho(B)}}{\rho(B)!} ,
\end{align}
with $\rho(B)!$ denoting the factorial of $\rho(B)$. Under certain assumptions, it can be shown that the stochastic model converges in probability to the deterministic model, as the molecule counts tend to infinity. These are commonly known as fluid limit results~\cite{KurtzBible,DBLP:journals/pe/BortolussiHLM13}, as discussed in Section~\ref{sec:ucrn:lump:fluid}.

We now introduce the notion of \emph{controllable CRN} (CCRN), for which we likewise give both a stochastic and a deterministic control system. In both cases we consider two extremal CRNs $(\calS, \calR_{\lb})$ and $(\calS, \calR_{\ub})$, with $\lb \leq \ub$, which constrain the values that the decision variables (i.e., the control inputs) may attain.
\begin{itemize}
\item The stochastic control system is given by~(\ref{eq:forward}), where each $q(\sigma,\theta)$ becomes a measurable control input bounded by the corresponding values in the extremal CRNs, that is
    \begin{align}\label{eq:q:var}
    q(\sigma,\theta) = q_{\lb,\ub}(\sigma,\theta) : \RE_{\geq0} \to [q_{\lb}(\sigma,\theta); q_{\ub}(\sigma,\theta)]
    \end{align}
    The resulting family of CTMCs is denoted by $\big(\NA^\calS,[ q_{\lb} ; q_{\ub}]\big)$ and is called the \emph{uncertain CTMC} (UCTMC) of a CCRN.\footnote{We use here the name from~\cite{DBLP:conf/qest/CardelliGLTTV21}, even though the name \emph{controlled CTMC} would be appropriate too.}

\item Likewise, in the deterministic control system, each kinetic parameter in~(\ref{eq:det}) becomes a control input bounded by the corresponding values in the extremal CRNs, that is a measurable $\alpha_{i_r} : \RE_{\geq 0} \rightarrow [\lb_{i_r} ; \ub_{i_r}]$. Moreover, for any bounded set of initial conditions $I \subseteq \RE_{\geq0}^\calS$ and time $t \geq 0$, we define the set of states reachable from $I$ at time $t$ as
    \[
    \frR(t) = \{  v(t) \mid \partial_t{v} = f(v,\alpha), v(0) \in I, \alpha \in [\lb;\ub] \}
    \]
    The initial set $I$ allows to account for uncertainty in the initial condition and encapsulates as special case the singleton set. For a given $\alpha : \RE_{\geq 0} \rightarrow [\lb ; \ub]$, we shall write $v^\alpha$ for the solution of $\partial_t{v} = f(v,\alpha)$, where $v(0) \in I$ is assumed to be given.
\end{itemize}

We shall adhere to the following notation.

\begin{remark}
Since $q$ is $q_{\lb,\ub}$ from~(\ref{eq:q:var}) rather than $q_\alpha$ from~(\ref{eq:q:const}) in all but few cases, $q$ shall refer to $q_{\lb,\ub}$ unless stated otherwise. Also, we shall write $\alpha_i$ rather than $\alpha_{i_r}$ to increase readability.
\end{remark}

%

%
%
%

We end the section by pointing out the following.

\begin{remark}
In general, ensuring that the forward equation~(\ref{eq:forward}) is regular in the sense that it admits a unique solution for every initial probability distribution $p(0)$ is nontrivial because the state space $\NA^\calS$ is infinite. A common way to ensure regularity is to prove that the CTMC is non-explosive by means of stochastic Lyapunov conditions~\cite{10.2307/1427522,EberleLecture08}.
\end{remark}

We call a CCRN/UCTMC regular if it induces regular CTMCs only. While some of our results assume regularity, the optimality-preserving lumping from Section~\ref{sec:ucrn:lump:fluid} does not.

\section{Species Equivalence of controlled CRNs}\label{sec:ucrn:lump}

We shall use the following CCRN as running example.

\begin{example}\label{ex:model}
Consider the CCRN $(\calS,\calR_{[\lb ; \ub]})$ with species $\calS = \{B, A_{00}, A_{01}, A_{10}, A_{11} \}$ and reactions
\begin{align*}
& A_{00} + B  \act{[\lb_1;\ub_1]} A_{10}, &
& A_{10} \act{[\lb_2;\ub_2]} A_{00} + B, \\
& A_{00} + B  \act{[\lb_3;\ub_3]} A_{01}, &
& A_{01} \act{[\lb_4;\ub_4]} A_{00} + B, \\
& A_{10} + B \act{[\lb_5;\ub_5]} A_{11}, &
& A_{11} \act{[\lb_6;\ub_6]} A_{10} + B, \\
& A_{01} + B \act{[\lb_7;\ub_7]} A_{11}, &
& A_{11} \act{[\lb_8;\ub_8]} A_{01} + B.
\end{align*}
The reactions model reversible binding of species $B$ to a substrate $A$ with two binding sites. Subscripts $i,j$ in chemical species $A_{ij}$ denote the availability of either binding site in the substrate $A$, while the value on each arrow indicates the kinetic rate parameter. For state $\sigma = A_{01} + A_{10} + B$, these yield
\begin{align*}
q_{\sigma,A_{11} + A_{10}}(\cdot) & \in [\lb_7;\ub_7], &
q_{\sigma,A_{00} + A_{10} + 2B}(\cdot) & \in [\lb_4;\ub_4], \\
q_{\sigma,A_{11} + A_{01}}(\cdot) & \in [\lb_5;\ub_5], &
q_{\sigma,A_{00} + A_{01} + 2B}(\cdot) & \in [\lb_2;\ub_2] .
\end{align*}
In the following, we make the common assumption~\cite{PNAScttv} that the uncertainty intervals do not depend on the binding site, that is, $[\lb_i;\ub_i] = [\lb_{i+1};\ub_{i+1}]$ for $i \in \{1,2,5,6\}$.
\end{example}

\subsection{Lumping of CCRNs}

Ordinary lumpability is a partition $\calH$ of the state space such that any two states $i$, $j$ in each partition block $H \in \calH$ have equal aggregate rates toward states in any block $H' \in \calH$. That is, writing $\mathfrak{q}$ for the transition rates of a generic CTMC that is not necessarily related to~(\ref{eq:forward}), it must hold that $\sum_{k \in H'} \mathfrak{q}_{i,k}(\cdot) = \sum_{k \in H'} \mathfrak{q}_{j,k}(\cdot)$. Given an ordinarily lumpable partition, a lumped CTMC can be constructed by associating a macro-state to each block. Transitions between macro-states are labeled with the overall rate from a state in the source block toward all states in the target block.

Checking the conditions for ordinary lumpability requires the full enumeration of the CTMC state space which grows combinatorially in the multiplicities of the initial state and may be even infinite in presence of species creation (e.g., $A \act{\alpha} A + A$). Species equivalence~\cite{DBLP:journals/bioinformatics/CardelliPTTVW21} addresses this by detecting ordinary lumpability at the level of the reaction network. To this end, it identifies an equivalence relation which induces an ordinary lumpable partition over the multisets representing CTMC states. Specifically, one considers a natural lifting of a partition $\calH$ of species to multisets of species, called \emph{multiset lifting} of $\calH$, denoted by $\upp{\pa}$.




\begin{definition}[Multiset Lifting]\label{def:msLifting} 
Let $(\calS,\calR_{[\lb;\ub]})$ be a CCRN, a partition $\calH$ over $\calS$ and let $\calE$ be the equivalence relation of $\calH$, i.e., $\calH = \calS/\calE$. We define the multiset lifting of $\calE$ on $\NA^\calS$, denoted by $\upp{\calE} \subseteq \NA^\calS \times \NA^\calS$, as
\[
\big\{ (\sigma_1,\sigma_2) \in \NA^\calS \times \NA^\calS \mid \forall H \in \calH . \sum_{A \in H}\sigma_1(A) = \sum_{A \in H}\sigma_2(A) \big\}
\]
With this, we set $\upp{\calH} = \NA^\calS / \upp{\calE}$.
\end{definition}

Intuitively, the multiset lifting relates multisets that have same cumulative multiplicity from each partition block.

\begin{example}\label{ex:lifting}
In Example~\ref{ex:model}, consider $\calH = \{\{B\}, \{A_{00}\},$ $\{A_{01}, A_{10}\},\{ A_{11} \}\}$ and let $\calE$ be such that $\calH = \calS/\calE$. Then, $(A_{01},A_{10}) \in \upp{\calE}$, $(2A_{01},A_{01}+A_{10}) \in \upp{\calE}$, while $(A_{00},A_{10}) \not\in \upp{\calE}$ and $(2A_{01},A_{10}) \notin \upp{\calE}$. That is, two species are equivalent w.r.t. $\upp{\calE}$ when they agree on the number of occupied binding sites. More formally, $(\sigma,\sigma') \in \upp{\calE}$ whenever $\sigma(C) = \sigma'(C)$ for all $C \not \in \{A_{00}, A_{11}\}$ and $\sigma(A_{01}) + \sigma(A_{10}) = \sigma'(A_{01}) + \sigma'(A_{10})$.
\end{example}

We first review the notion of CRN species equivalence from~\cite{DBLP:journals/bioinformatics/CardelliPTTVW21}.

\begin{definition}[CRN Species Equivalence]\label{def:ucrn:lump}
Fix a CRN $(\calS,\calR_\alpha)$. We call a partition $\calH$ of $\calS$ a CRN species equivalence if, for any two species $A_i,A_j$ in a block of $\calH$, any reagent $\rho \in \mathbb{N}_0^\calS$, any block $\upp{H} \in \upp{\calH}$, we have
\begin{equation}\label{eq:rr}
\sum_{\pi \in \upp{H}} \rr_\alpha(A_i + \rho, \pi) =  \sum_{\pi \in \upp{H}} \rr_\alpha(A_j + \rho, \pi)
\end{equation}
Here, $\rr_\alpha$ is the reaction rate from $\rho$ to $\pi$
\[
\rr_\alpha(\rho,\pi) =
\begin{cases}
\sum\limits_{{(\rho \act{\alpha_i} \pi) \in \calR_\alpha}} \alpha_i & , \ \rho \neq\pi \\
-\sum\limits_{\pi' \neq \rho} \rr_\alpha(\rho,\pi') & , \ \rho = \pi
\end{cases}
\]
For any $\upp{H} \subseteq \NA^\calS$, we set $\rr_\alpha[\rho,\upp{H}] = \sum_{\pi\in \upp{H}} \rr_\alpha(\rho,\pi)$.
\end{definition}

Any CRN species equivalence induces a lumped CRN given next.

\begin{definition}[Lumped CRN]\label{def:reduced:CRN}
Let $(\calS,\calR_\alpha)$ be a CRN, $\calH$ a CRN species equivalence and fix a representative $A_H \in H$ for each $H \in \calH$. The lumped CRN is then given by $(\hS,\hR_{\hat{\alpha}})$, where the species are $\hS = \{ A_H \mid H \in \calH \}$, while reactions $\hR_{\hat{\alpha}}$ arise via
\begin{enumerate}
    \item discard all reactions $\rho \act{\alpha_i} \pi$ where $\rho$ has a nonrepresentative species;
    \item replace the species in the products of the remaining reactions by their representatives;
    \item fuse all reactions that have the same reactants and products by summing their rates.
\end{enumerate}
\end{definition}

With the foregoing definitions in place, CCRN species equivalence is defined as the CRN species equivalence of the extremal CRNs $(\calS,\calR_{\lb})$ and $(\calS,\calR_{\ub})$.

\begin{definition}[CCRN Species Equivalence]\label{def:ucrn:lump}
Fix a CCRN $(\calS,\calR_{[\lb;\ub]})$. We call a partition $\calH$ of $\calS$ a CCRN species equivalence whenever $\calH$ is a CRN species equivalence of $(\calS,\calR_{\lb})$ and $(\calS,\calR_{\ub})$.
\end{definition}

Our example enjoys a CCRN species equivalence.

\begin{example}\label{ex:ucrn:lump3}
Continuing Example~\ref{ex:model} and~\ref{ex:lifting}, we note that for
\[
\alpha = (\alpha_1,\ldots,\alpha_8) \in \{(\lb_1,\ldots,\lb_8),(\ub_1,\ldots,\ub_8)\} ,
\]
it holds that $\rr_\alpha(A_{01},A_{00}+B)$ $=\rr_\alpha(A_{10},A_{00}+B)$ and $\rr_\alpha(A_{01} + B,A_{11})$ $=\rr_\alpha(A_{10} + B,A_{11})$. Hence, $\calH$ is a CCRN species equivalence.
\end{example}

The lumped CCRN is given by the lumpings of the extremals, as stated next.

\begin{definition}[Lumped CCRN]\label{def:reduced:CCRN}
Let $(\calS,\calR_{[\lb;\ub]})$ be a CCRN and $\calH$ a CCRN species equivalence. The lumped CCRN $(\hS,\hR_{[\hat{\lb};\hat{\ub}]})$ arises by lumping the extremal CRNs $(\calS,\calR_{\lb})$ and $(\calS,\calR_{\ub})$ as outlined in Definition~\ref{def:reduced:CRN}.
\end{definition}

We remark that the lumped CCRN does not depend on the choice of the representative~\cite{DBLP:journals/bioinformatics/CardelliPTTVW21}. As next, we provide the lumped CCRN of our example.

\begin{example}\label{ex:ucrn:lump4}
Continuing Example~\ref{ex:model}-\ref{ex:ucrn:lump3}, the lumped CCRN is given by $\hS = \calS \setminus \{ A_{10} \}$ and $\hR_{[\hat{\lb};\hat{\ub}]}$ such that
\begin{align*}
& A_{00} + B  \act{[\lb_1+\lb_3;\ub_1+\ub_3]} A_{01}, &
& A_{01} \act{[\lb_4;\ub_4]} A_{00} + B, \\
& A_{11} \act{[\lb_6+\lb_8;\ub_6+\ub_8]} A_{01} + B, &
& A_{01} + B  \act{[\lb_7;\ub_7]} A_{11}
\end{align*}
\end{example}

\begin{algorithm}[tp!]
\caption{Algorithm for the computation of the coarsest CCRN species equivalence that refines some given partition $\calG$.}\label{algorithm_part}
\begin{algorithmic}
\REQUIRE CCRN $(\calS,\calR_{[\lb;\ub]})$ and a partition $\calG$ of $\calS$
\WHILE{\TRUE}
    \STATE \textbf{compute} using the algorithm in~\cite{DBLP:journals/bioinformatics/CardelliPTTVW21} the coarsest CCRN species equivalence of $(\calS,\calR_{\lb})$ that refines $\calG$, \textbf{store} it in $\calH$
    \STATE \textbf{compute} using the algorithm in~\cite{DBLP:journals/bioinformatics/CardelliPTTVW21} the coarsest CCRN species equivalence of $(\calS,\calR_{\ub})$ that refines $\calH$, \textbf{store} it in $\calH'$
    \IF{$\calH' = \calG$}
        \RETURN $\calG$
    \ELSE
        \STATE $\calG \leftarrow \calH'$
    \ENDIF
\ENDWHILE
\end{algorithmic}
\end{algorithm}

The CCRN species equivalence can be computed by invoking alternately the CRN lumping algorithm from~\cite{DBLP:journals/bioinformatics/CardelliPTTVW21} on the extremal CRNs that define a CCRN, as stated next.

\begin{theorem}[Computation of CCRN Species Equivalence]\label{thm:ucrn:lump}
Let $(\calS,\calR_{[\lb;\ub]})$ be a CCRN. Then we have the following.
\begin{enumerate}
    \item $\calH$ is a CCRN species equivalence iff $\upp{\calH}$ is an ordinary lumpability of the CTMCs $(\NA^\calS,q_{\lb})$ and $(\NA^\calS,q_{\ub})$.

    \item For any partition $\calG$ of $\calS$, Algorithm~\ref{algorithm_part} computes the coarsest CCRN species equivalence of $(\calS,\calR_{[\lb;\ub]})$ that refines $\calG$. That is, $\calH$ is such that
        \begin{itemize}
            \item for every block $G \in \calG$, there exist unique blocks $H_1,\ldots,H_l \in \calH$ such that $G = H_1 \cup \ldots \cup H_l$ and;
            \item $\calH$ is a CCRN species equivalence and has a minimal number of blocks, hence a lumped CCRN of minimal size.
        \end{itemize}
        The number of steps performed by Algorithm~\ref{algorithm_part} is polynomial in $|\calS|$ and $|\calR_{\lb}|$.
\end{enumerate}
\end{theorem}

\begin{proof}
We start by noting that $\calH$ is a CCRN species equivalence of $(\calS,\calR_{[\lb;\ub]})$ if and only if $\calH$ is a CRN species equivalence~\cite{DBLP:journals/bioinformatics/CardelliPTTVW21} of $(\calS,\calR_{\lb})$ and $(\calS,\calR_{\ub})$. With this in mind, we first observe that 1) and 2) follow directly from~\cite{DBLP:journals/bioinformatics/CardelliPTTVW21} in the special case $\lb = \ub$. Let us now consider the general case $\lb < \ub$. Then, 1) follows from the special case of 1) and the definition of CCRN species equivalence. Likewise, 2) follows by the definition of Algorithm~\ref{algorithm_part} and the special case of 2).
\end{proof}

After addressing the computation of CCRN species equivalence, we observe next that the block sums of original CCRN states are equivalent in distribution to the states of the lumped CCRN. Following standard notation, equivalence in distribution is denoted by $\deq$.

\begin{theorem}[CCRN Species Equivalence]\label{thm:ucrn:lump:2}
Let $\calH$ be a CCRN species equivalence of $(\calS,\calR_{[\lb;\ub]})$ and let $(\hS,\hR_{[\hat{\lb};\hat{\ub}]})$ be the respective lumped CCRN. Moreover, let $X^q$ and $\hX^{\hq}$ denote family members of the respective UCTMCs. Then, if both CCRNs are regular, we have:
\begin{enumerate}
    \item For any $q(\cdot) \in [ q_{\lb} ; q_{\ub}]$, there exists a  $\hq(\cdot) \in [ q_{\hat{\lb}} ; q_{\hat{\ub}}]$ such that
        \begin{align}\label{eq:uctm:lump:2}
        \sum_{A \in H} X^q_A(t) \deq \hX^{\hq}_{A_H}(t) , \quad \forall H \, \in \calH \forall \, t > 0 ,
        \end{align}
        provided the statement holds for $t = 0$.
    \item Conversely, for any  $\hq(\cdot) \in [q_{\hat{\lb}} ; q_{\hat{\ub}}]$, there is a  $q(\cdot) \in [ q_{\lb} ; q_{\ub}]$ with~(\ref{eq:uctm:lump:2}).
\end{enumerate}
\end{theorem}

\begin{proof}
Let us first assume that $\lb = \ub$. Then, for any block $\upp{H} \in \upp{\calH}$ and its representative $\sigma_{\upp{H}} \! \in \! \upp{H} \cap \NA^{\hS}$, statement 1) of Theorem~\ref{thm:ucrn:lump} and the regularity ensure~\cite{DBLP:journals/bioinformatics/CardelliPTTVW21} that
\begin{align*}
\sum_{\sigma \in \upp{H}} p_\sigma(t) = \hat{p}_{\sigma_{\upp{H}}}(t) , \quad \forall \, t > 0 .
\end{align*}
Together with
\[
\Big\{ \sigma \in \NA^{\calS} \mid \forall H \in \calH . \sum_{A \in H} \sigma_A = (\sigma_{\upp{H}})_{A_H} \Big\} = \upp{H} ,
\]
this implies $\big(\sum_{A \in H} X_A(t)\big)_{H \in \calH} \deq \big(\hX_{A_H}(t)\big)_{H \in \calH}$, yielding the claim. The general case, instead, follows from the proof of Theorem 6 from~\cite{DBLP:conf/qest/CardelliGLTTV21}. Specifically, the proof carries over verbatim to our setting of countable state spaces because $\upp{\calH}$ has finite blocks and the assumption of regularity ensures that the forward Kolmogorov equations enjoy unique solutions.
\end{proof}

Provided the $n$-th order moments exist, Theorem~\ref{thm:ucrn:lump:2} implies in particular $\mathbb{E}\big[ (\sum_{A \in H} X_A(t))^n \big] = \mathbb{E}\big[\hX^n_{A_H}(t)\big]$.\footnote{Similarly to regularity, the existence of moments can be addressed by means of Lyapunov conditions~\cite{10.2307/1427522,EberleLecture08}} The moments, in turn, can be estimated by means of stochastic simulation~\cite{doi:10.1063/1.2799998}.

\begin{example}\label{ex:thm:ucrn:lump}
It can be shown that Example~\ref{ex:model} is regular. With this, Theorem~\ref{thm:ucrn:lump:2} essentially ensures that
\begin{itemize}
    \item for any $q$, there exists a $\hq$ such that $X^q_{A_{01}} + X^q_{A_{10}} \deq \hX^{\hq}_{A_{01}}$ and $X^q_S \deq \hX^{\hq}_S$ with $S \notin \{A_{01},A_{10}\}$;
    \item for any $\hq$, there exists a $q$ such that $X^q_{A_{01}} + X^q_{A_{10}} \deq \hX^{\hq}_{A_{01}}$ and $X^q_S \deq \hX^{\hq}_S$ with $S \notin \{A_{01},A_{10}\}$.
\end{itemize}
That is, if one is only interested in species $S \notin \{A_{01}, A_{10}\}$ or the cumulative behavior of species $X^q_{A_{01}} + X^q_{A_{10}}$, any behavior of the original CCRN can be matched by the lumped CCRN and vice versa.
\end{example}

Example~\ref{ex:thm:ucrn:lump} demonstrates that CRN species equivalence allows one to lump the original CCRN to a smaller lumped CCRN at the expense of preserving sums of original species. For instance, if  the modeler is interested in $X_{A_{00}}$ and $X_B$, partition $\calH$ from Example~\ref{ex:lifting} can be used because $\{A_{00}\}, \{B\} \in \calH$. Instead, if a modeler is interested in $X_{A_{10}}$, it is not possible to use $\calH$ because the lumped CCRN would only capture the cumulative behavior $X_{A_{10}} + X_{A_{01}}$. A natural question would be then if there is a CCRN species equivalence $\calH'$ which contains $\{A_{10}\}$. This can be readily checked by applying the algorithm from Theorem~\ref{thm:ucrn:lump} to $\calG' = \{ \{ A_{10} \} , \calS \setminus \{ A_{10} \} \}$. This is because any CCRN species equivalence $\calH'$ refining $\calG'$ has to contain the block $\{ A_{10} \}$. An application of the algorithm returns then the trivial CCRN species equivalence $\calH' = \{ \{ S \} \mid S \in \calS \}$. We call $\calH'$ trivial because it does not lump the original CCRN.

\section{Optimality-preserving Lumping}\label{sec:ucrn:lump:fluid}


We start by providing the deterministic control system of our example.

\begin{example}\label{ex:fmodel}
The CCRN $(\calS,\calR_{[\lb;\ub]})$ from Example~\ref{ex:model} gives rise to the ODE system
\begin{align}\label{eq:fmodel}
\partial_t v_{A_{00}} & = -(\alpha_1 + \alpha_3) v_{A_{00}} v_B + \alpha_2 v_{A_{10}} + \alpha_4 v_{A_{01}} \nonumber \\
\partial_t v_{A_{10}} & = \alpha_1 v_{A_{00}} v_B - \alpha_2 v_{A_{10}} - \alpha_5 v_{A_{10}} v_B + \alpha_6 v_{A_{11}} \nonumber \\
\partial_t v_{A_{01}} & = \alpha_3 v_{A_{00}} v_B - \alpha_4 v_{A_{01}} - \alpha_7 v_{A_{01}} v_B + \alpha_8 v_{A_{11}} \\
\partial_t v_{A_{11}} & = \alpha_5 v_{A_{10}} v_B + \alpha_7 v_{A_{01}} v_B - (\alpha_6 + \alpha_8) v_{A_{11}} \nonumber \\
\partial_t v_{B} & = - (\alpha_1 + \alpha_3) v_{A_{00}} v_B + \alpha_2 v_{A_{10}} + \alpha_4 v_{A_{01}} \nonumber \\
& \qquad - \alpha_5 v_{A_{10}} v_B - \alpha_7 v_{A_{01}} v_B + (\alpha_6 + \alpha_8) v_{A_{11}} \nonumber
\end{align}
\end{example}


The deterministic control system~(\ref{eq:det}) is also known as the fluid model of a CRN. This is because it can be approximated by CTMCs that have as states, loosely speaking, fractions $\frac{1}{N} \NA^\calS$ rather than integers $\NA^\calS$~\cite{kurtz-chem,dsn16BortolussiGast}.

\begin{definition}[CTMC Approximation]\label{def:ctmc:approx}
Fix a CRN $(\calS,\calR_\alpha)$ and a constant $c > 0$. The $N$-th CTMC approximation of $(\calS,\calR_\alpha)$ is $X_N = (\frac{1}{N} \NA^\calS,q^N_{\alpha})$, where for two different $\sigma, \theta \in \frac{1}{N} \NA^\calS$ we have:
\begin{align*}
g(\sigma) & = \max\{0,\min\{1,2 - | \sigma | / c \}\} \\
q^N_\alpha(\sigma,\theta) & = g(\sigma) \cdot q_{\alpha^N}(N\sigma,N\theta) ,
\end{align*}
where each $\rho \! \act{\alpha_i} \pi \! \in \! \calR_\alpha$ induces a $\rho \! \act{\alpha^N_i} \pi \! \in \! \calR_{\alpha^N}$ with $\alpha^N_i = \alpha_i / N^{|\rho|-1}$ for $|\rho| = \sum_{A \in \calS} \rho(A)$.\footnote{The CTMC approximation could be given without a cutoff function $g$. It will be mainly needed for the UCTMC counterpart from Definition~\ref{def:uctmc:approx}.}
\end{definition}

Generalizing the foregoing notion, we introduce a UCTMC approximation of a CCRN.

\begin{definition}[UCTMC Approximation]\label{def:uctmc:approx}
Fix a CCRN $(\calS,\calR_{[\lb;\ub]})$ and a constant $c > 0$. The $N$-th UCTMC approximation of $(\calS,\calR_{[\lb;\ub]})$ is $X_N = (\frac{1}{N} \NA^\calS,q^N_{[\lb;\ub]})$, where $q^N_{[\lb;\ub]} = [q^N_{\lb} ; q^N_{\ub}]$, with $q^N_{\lb}$ and $q^N_{\ub}$ as in Definition~\ref{def:ctmc:approx}.
\end{definition}

We next prove that the UCTMCs $X_N$ converge to the fluid CCRN model of $(\calS,\calR_{[\lb;\ub]})$. To this end, we first show that for any $\alpha(\cdot) \in [\lb ; \ub]$ there exists a $q(\cdot) \in q^N_{[\lb;\ub]}$ such that the ODE solution $v^{\alpha}$ is sufficiently close to the CTMC simulation $X^q_N$, provided that $N$ is large enough and $X^q_N$ denotes the CTMC induced by $q$. This follows from standard fluid limit results~\cite[{\S}11.1-{\S}11.2]{KurtzBible}.

\begin{proposition}\label{prop:limit:0}
Fix a CCRN $(\calS,\calR_{[\lb;\ub]})$, a time $T > 0$ and assume that $\frR(t) \subseteq B(c)$ for all $t \in [0;T]$, where $B(c)$ is the $L_1$ ball with radius $c$ centered at the origin. Assume further that the UCTMC approximations $(X_N(0))_{N \geq 1}$ satisfy $X_N(0) = \lfloor N v(0) \rfloor / N$ for some $v(0) \in I$ in the set of initial conditions $I$ of the CCRN. Then, for any $\varepsilon, \delta > 0$, there exists an $N \geq 1$ such that for any $\alpha(\cdot) \in [\lb ; \ub]$, there exists a $q(\cdot) \in q^N_{[\lb;\ub]}$ such that
\[
\mathbb{P}\big\{ \sup_{0 \leq t \leq T} | X^q_N(t) - v^\alpha(t) | > \varepsilon \big\} < \delta .
\]
\end{proposition}

\begin{proof}
We use~\cite[{\S}11.2, Theorem 2.1]{KurtzBible}. Specifically, we first note that the discussion in~\cite[{\S}11.1-{\S}11.2]{KurtzBible} readily extends to time-varying transition rates. Moreover, it is possible to use uniform estimations in the proof of Theorem 2.1 which do not depend on the choice of $\alpha(\cdot) \in [\lb ; \ub]$. Armed with this insight, we pick any $\alpha(\cdot) \in [\lb ; \ub]$ and consider the CTMCs $Z_N = (\frac{1}{N} \NA^\calS,\mathfrak{q}^N_{\alpha(\cdot)})$, where
\begin{equation}\label{eq:q:classic}
\mathfrak{q}^N_{\alpha(\cdot)}(\sigma,\theta) =
\sum_{ \substack{ r = (\rho \act{\alpha_i(\cdot)} \pi ) \in \calR_{\alpha(\cdot)} \\ \theta = \sigma + \frac{1}{N}(\pi - \rho) } } \frac{\alpha_i(\cdot)}{ N^{|\rho|-1} } \cdot \binom{N \sigma}{\rho} 
\end{equation}
The result then follows by applying Theorem 2.1 to $\min\{T,\tau\}$ rather than $T$, where $\tau$ is the first exit time of $\{ \sigma \in \frac{1}{N} \NA^\calS \mid | \sigma | \leq c \}$, see also~\cite[Corollary 2.8]{10.2307/3212147}. Crucially, due to uniform estimations, one can pick an $N$ such that the statement holds for all $\alpha(\cdot) \in [\lb ; \ub]$.
\end{proof}

Our second approximation result ensures, conversely, that for any $q(\cdot) \in q^N_{[\lb;\ub]}$ there exists a $\alpha(\cdot) \in [\lb ; \ub]$ such that the ODE solution $v^{\alpha}$ is sufficiently close to the CTMC simulation $X^q_N$, provided that $N$ is large enough.

\begin{proposition}\label{prop:limit}
Under the same assumptions as Proposition~\ref{prop:limit:0} and for any $\varepsilon, \delta > 0$, there exists an $N \geq 1$ such that for any $q(\cdot) \in q^N_{[\lb;\ub]}$, there exists $\alpha(\cdot) \in [\lb ; \ub]$ such that
\[
\mathbb{P}\big\{ \sup_{0 \leq t \leq T} | X^q_N(t) - v^\alpha(t) | > \varepsilon \big\} < \delta .
\]
\end{proposition}

\begin{proof}
To increase readability, we postpone the lengthy proof to Section~\ref{sec:postponed}.
\end{proof}

\subsection{Proof of Optimality-Preservation}

Before proving the main result, we establish our last auxiliary result which ensures that the transient probabilities of the $N$-th UCTMC approximation of the original and the lumped CCRN coincide on the blocks of $\upp{H}$, if $N$ is large enough and $\calH$ is a CCRN species equivalence. The proof relies on~\cite{DBLP:conf/qest/CardelliGLTTV21}.

\begin{proposition}\label{prop:lumped:uctmc:approx}
Let $\calH$ be a CCRN species equivalence of $(\calS,\calR_{[\lb;\ub]})$ and let $X_N$ and $\hX_N$ denote, respectively, the $N$-UCTMC approximation of the original and the lumped CCRN, see Definition~\ref{def:reduced:CCRN} and~\ref{def:uctmc:approx}. Then, we have the following.
\begin{itemize}
    \item For any $q(\cdot) \in q^N_{[\hat{\lb};\hat{\ub}]}$ there is a $\hq(\cdot) \in q^N_{[\hat{\lb};\hat{\ub}]}$ such that
    \begin{align}\label{eq:lumped:uctmc}
        \forall \upp{H} \in \upp{\calH} . \, \sum_{\sigma \in \upp{H}} p_{\frac{1}{N} \sigma}(t) = \hp_{\frac{1}{N} \sigma_{\upp{H}}}(t)
    \end{align}
    holds for all $t > 0$, provided it holds for $t = 0$. Here, $p$ and $\hp$ is the transient probability of $X_N^q$ and $\hX_N^{\hq}$, respectively, while $\sigma_{\upp{H}} \in \upp{H} \cap \NA^{\hS}$ is the unique representative of $\upp{H}$. 
    \item Conversely, for any $\hq(\cdot) \in q^N_{[\hat{\lb};\hat{\ub}]}$ there is a $q(\cdot) \in q^N_{[\hat{\lb};\hat{\ub}]}$ such that~(\ref{eq:lumped:uctmc}) holds for all $t > 0$, if it holds for $t = 0$.
\end{itemize}
\end{proposition}

\begin{proof}
We begin by proving that $\calH$ is a CCRN species equivalence of $(\calS,\calR_{[\lb^N;\ub^N]})$, where $\lb^N$ and $\ub^N$ are as in Definition~\ref{def:ctmc:approx}. This holds true if $\calH$ is a CRN species equivalence of $(\calS,\calR_{\alpha^N})$ with $\alpha^N \in \{\lb^N, \ub^N\}$. To see this, pick any $\alpha \in \{\lb,\ub\}$, $H \in \calH$, $A_i, A_j \in H$ and $\upp{H} \in \upp{\calH}$. Then, we need to show that
\begin{equation}\label{rrN=rrN}
\sum_{\pi \in \upp{H}} \rr^N_{\alpha}(A_i + \rho, \pi) =  \sum_{\pi \in \upp{H}} \rr^N_{\alpha}(A_j + \rho, \pi) .
\end{equation}
Here, $\rr^N_{\alpha}$ is defined according to Definition~\ref{def:ucrn:lump} as
\begin{multline*}
\rr^N_{\alpha}(A_k + \rho,\pi) = \\
\begin{cases}
\sum\limits_{{(A_k + \rho \act{\alpha^N_i} \pi) \in \calR_{\alpha^N}}} \frac{\alpha_i}{N^{|A_k + \rho|-1}} & , \  A_k + \rho \neq\pi, \\
-\sum\limits_{\pi' \neq A_k + \rho} \rr^N_{\alpha}(A_k + \rho,\pi') & , \ A_k + \rho = \pi .
\end{cases}
\end{multline*}
Then obviously $\rr^N_{\alpha}(A_k + \rho,\pi) = \frac{\rr_\alpha(A_k + \rho,\pi)}{N^{|A_k + \rho|-1}}$. As $\calH$ is a CRN species equivalence of $(\calS,\calR_\alpha)$ by assumption, it holds that
\begin{equation*}
\sum_{\pi \in \upp{H}} \rr_\alpha(A_i + \rho, \pi) =  \sum_{\pi \in \upp{H}} \rr_\alpha(A_j + \rho, \pi) ,
\end{equation*}
thus showing that $\calH$ is indeed a CCRN species equivalence of $(\calS,\calR_{[\lb^N;\ub^N]})$. Noting that $g(\frac{\sigma}{N}) = g(\frac{\sigma'}{N})$ for all $\sigma, \sigma' \in \upp{H}$ and $\upp{H} \in \upp{\calH}$, this implies that $\frac{1}{N} \upp{\calH}$ is a CTMC lumpability of CTMCs $(\frac{1}{N} \NA^\calS,q^N_{\lb})$ and $(\frac{1}{N} \NA^\calS,q^N_{\ub})$. Since $X_N$ and $\hX_N$ are regular due to function $g$, the statement follows by arguing as in the proof of Theorem~\ref{thm:ucrn:lump:2}.
\end{proof}

With Proposition~\ref{prop:limit:0}-\ref{prop:lumped:uctmc:approx}, we are in a position to state and prove that CCRN species equivalence preserves the deterministic models. This, in turn, will be key in proving the preservation of optimality. The proof strategy is visualized in Fig.~\ref{fig:proof}.

\begin{theorem}[Deterministic CCRN Lumping]\label{thm:ucrn:lump:fluid}
Let us fix a CCRN $(\calS,\calR_{[\lb;\ub]})$, a constant $c > 0$, assume that $\calH$ is a CCRN species equivalence and denote the corresponding lumped CCRN by $(\hS,\hR_{[\hat{\lb};\hat{\ub}]})$. If $T > 0$ is such that $\frR(t) \subseteq B(c)$ for any $t \in [0;T]$, then for any initial condition $v(0) \in I$ and any $\varepsilon, \delta > 0$, the original and lumped deterministic models, $\partial_t v^\alpha = f(v^\alpha,v)$ and $\partial_t \hv^{\ha} = \hf(\hv^{\ha},\hv)$, enjoy the following.
\begin{enumerate}
    \item For any $\alpha(\cdot) \in [\lb;\ub]$, there is some $\ha(\cdot) \in [\hat{\lb};\hat{\ub}]$ such that $\partial_t v^\alpha = f(v^\alpha,v)$ and $\partial_t \hv^{\ha} = \hf(\hv^{\ha},\hv)$ satisfy
        \[
        \mathbb{P}\Big\{ \max_{H \in \calH} \max_{0 \leq t \leq T} \big|\sum_{A \in H} v^\alpha_A(t) - \hv^{\ha}_{A_H}(t)\big| > \varepsilon \Big\} < \delta
        \]
        provided that $\sum_{A \in \calH} v_A(0) = \hv_{A_H}(0)$ for all $H \in \calH$.
    \item For any $\ha(\cdot) \in [\hat{\lb};\hat{\ub}]$, there is some $\alpha(\cdot) \in [\lb;\ub]$ such that $\partial_t v^\alpha = f(v^\alpha,v)$ and $\partial_t \hv^{\ha} = \hf(\hv^{\ha},\hv)$ satisfy
        \[
        \mathbb{P}\Big\{ \max_{H \in \calH} \max_{0 \leq t \leq T} \big|\sum_{A \in H} v^\alpha_A(t) - \hv^{\ha}_{A_H}(t)\big| > \varepsilon \Big\} < \delta
        \]
        provided that $\sum_{A \in \calH} v_A(0) = \hv_{A_H}(0)$ for all $H \in \calH$.
\end{enumerate}
\end{theorem}

\begin{figure}[tp]
    \begin{center}
        \begin{tikzpicture}[->,>=stealth',shorten >=1pt,auto,semithick]
        \matrix[row sep=1cm, column sep=1cm]
        {
        \pgfmatrixnextcell \node (1) {$\substack{\dsl \partial_t p = p^T q \\ \dsl (\calS,\calR_{[\lb^N;\ub^N]})}$};
        \pgfmatrixnextcell \node (2) {$\substack{\dsl \partial_t \hp = \hp^T \hq \\ \dsl (\hS,\hR_{[\hat{\lb}^N;\hat{\ub}^N]}) }$}; \\
        \pgfmatrixnextcell \node (3) {$\substack{\dsl \partial_t v = f(v,\alpha) \\ \dsl (\calS,\calR_{[\lb;\ub]})}$};
        \pgfmatrixnextcell \node (4) {$\substack{\dsl \partial_t \hat{v} = \hat{f}(\hat{v},\hat{\alpha}) \\ \dsl (\hS,\hR_{[\hat{\lb};\hat{\ub}]})}$}; \\
        };
        \path (1) edge node {Prop. \ref{prop:lumped:uctmc:approx}} (2)
            (3) edge [dashed] node {Thm. \ref{thm:ucrn:lump:fluid}} (4)
            (3) edge node {$\substack{\dsl \text{Prop. }\ref{prop:limit:0} \\ \dsl N \to \infty}$} (1)
            (2) edge node {$\substack{\dsl \text{Prop. }\ref{prop:limit} \\ \dsl N \to \infty}$} (4);
        \end{tikzpicture}
    \end{center}
\caption{Proof strategy of Theorem~\ref{thm:ucrn:lump:fluid}, part 1). The result is proven by approximating the deterministic control systems of the original and the lumped CCRN by means of UCTMCs (Prop.~\ref{prop:limit:0} and~\ref{prop:limit}). This, in turn, are shown in Prop. 3 to coincide on the blocks (of the multiset lifting) of an ordinary lumpable partition. Part 2) is proven in a similar fashion by reversing the directions.}\label{fig:proof}
\end{figure}

\begin{proof}
We first prove 1). To this end, pick some small $\varepsilon, \delta > 0$ and some arbitrary $\alpha \in [\lb;\ub]$. By Proposition~\ref{prop:limit:0}-\ref{prop:limit}, we can pick an $N \geq 1$ such that
\begin{itemize}
    \item There is a $q \in q^N_{[\lb;\ub]}$ such that
    \[
    \mathbb{P}\big\{ \sup_{0 \leq t \leq T} | X^q_N(t) - v^\alpha(t) | > \varepsilon / |\calS| \big\} < \delta / |\calS| .
    \]
    \item For any $\hq \in q^N_{[\hat{\lb};\hat{\ub}]}$, there is an $\ha \in [\hat{\lb};\hat{\ub}]$ such that
    \[
    \mathbb{P}\big\{ \sup_{0 \leq t \leq T} | \hX^{\hq}_N(t) - \hv^{\ha}(t) | > \varepsilon \big\} < \delta .
    \]
\end{itemize}
Since $\calH$ is a CCRN species equivalence of $(\calS,\calR_{[\lb;\ub]})$, Proposition~\ref{prop:lumped:uctmc:approx} ensures that there is a $\hq \in q^N_{[\hat{\lb};\hat{\ub}]}$ such that the solutions of forward equations $\partial_t p^T = p^T q$ and $\partial_t \hp^T = \hp^T \hq$ satisfy
\[
\forall \upp{H} \in \upp{\calH} . \, \forall t \geq 0 . \, \sum_{\sigma \in \upp{H}} p_{\frac{1}{N} \sigma}(t) = \hp_{\frac{1}{N} \sigma_{\upp{H}}}(t)
\]
Moreover, for any $H \in \calH$, the first bullet point above and the inequality $\mathbb{P}\{|\sum_{i = 1}^\nu Z_i | > \eta \} \leq \sum_{i = 1}^\nu \mathbb{P}\{|Z_i| > \eta / \nu \}$, where $Z_i$ are real random variables, imply that
\[
\mathbb{P}\Big\{ \sup_{0 \leq t \leq T} \big| \sum_{A \in H} (X^q_N)_A(t) - \sum_{A \in H} v_A^\alpha(t) \big| > \varepsilon \Big\} < \delta .
\]
This and the foregoing choice of $\hq$ imply for all $H \in \calH$
\[
\mathbb{P}\Big\{ \sup_{0 \leq t \leq T} \big| (\hX^{\hq}_N)_{A_H}(t) - \sum_{A \in H} v_A^\alpha(t) \big| > \varepsilon \Big\} < \delta .
\]
Thanks to the second bullet point from above, we can pick an $\ha \in [\hat{\lb};\hat{\ub}]$ such that
\[
\mathbb{P}\big\{ \sup_{0 \leq t \leq T} | (\hX^{\hq}_N)_{A_H}(t) - \hv^{\ha}_{A_H}(t) | > \varepsilon \big\} < \delta .
\]
Using again $\mathbb{P}\{|\sum_{i = 1}^\nu Z_i | > \eta \} \leq \sum_{i = 1}^\nu \mathbb{P}\{|Z_i| > \eta / \nu \}$, the above discussion allows us thus to conclude that
\[
\mathbb{P}\big\{ \sup_{0 \leq t \leq T} | \sum_{A \in H} v_A^\alpha(t) - \hv_{A_H}^{\ha}(t) | > 2 \varepsilon \big\} < 2 \delta .
\]
Since the choice of $H \in \calH$ and $\varepsilon, \delta > 0$ was arbitrary, we obtain 1).

We prove 2) in a similar fashion. Specifically, thanks to Proposition~\ref{prop:limit:0}-\ref{prop:limit}, we can pick an $N \geq 1$ such that
\begin{itemize}
    \item There is a $\hq \in q^N_{[\hat{\lb};\hat{\ub}]}$ such that
    \[
    \mathbb{P}\big\{ \sup_{0 \leq t \leq T} | \hX^{\hq}_N(t) - \hv^{\ha}(t) | > \varepsilon \big\} < \delta .
    \]
    \item For any $q \in q^N_{[\lb;\ub]}$, there is an $\alpha \in [\lb;\ub]$ such that
    \[
    \mathbb{P}\big\{ \sup_{0 \leq t \leq T} | X^q_N(t) - v^\alpha(t) | > \varepsilon / |\calS| \big\} < \delta/|\calS| .
    \]
\end{itemize}
Using the first bullet point, we pick a $\hq$ such that for any $H \in \calH$ it holds
\[
\mathbb{P}\big\{ \sup_{0 \leq t \leq T} | \hX^{\hq}_{A_H}(t) - \hv^{\ha}_{A_H}(t) | > \varepsilon \big\} < \delta .
\]
Thanks to Proposition~\ref{prop:lumped:uctmc:approx}, we can further pick a $q \in q^N_{[\lb;\ub]}$ such that the solutions of forward equations $\partial_t p = p^T q$ and $\partial_t \hp = \hp^T \hq$ satisfy
\[
\forall \upp{H} \in \upp{\calH} . \, \forall t \geq 0 . \, \sum_{\sigma \in \upp{H}} p_{\frac{1}{N} \sigma}(t) = \hp_{\frac{1}{N} \sigma_{\upp{H}}}(t)
\]
Combining both statements yields
\[
\mathbb{P}\Big\{ \sup_{0 \leq t \leq T} \big| \sum_{A \in H} X^q_A(t) - \hv_{A_H}^{\ha}(t) \big| > \varepsilon \Big\} < \delta .
\]
Thanks to the second bullet point from above, we can pick next an $\alpha \in [\lb;\ub]$ such that
\[
\mathbb{P}\Big\{ \sup_{0 \leq t \leq T} \big| \sum_{A \in H} X^q_A(t) - \sum_{A \in H} v_A^\alpha(t) \big| > \varepsilon \Big\} < \delta
\]
The above discussion yields then
\[
\mathbb{P}\big\{ \sup_{0 \leq t \leq T} | \sum_{A \in H} v_A^\alpha(t) - \hv_{A_H}^{\ha}(t) | > 2 \varepsilon \big\} < 2 \delta .
\]
Since the choice of $H \in \calH$ and $\varepsilon, \delta > 0$ was arbitrary, we obtain 2).
\end{proof}


Let us next apply Theorem~\ref{thm:ucrn:lump:fluid} to our example.

\begin{example}\label{ex:fmodel:lumped}
The lumped CCRN $(\hS,\hR_{[\hat{\lb};\hat{\ub}]})$ from Example~\ref{ex:ucrn:lump4} has the fluid model
\begin{align}\label{eq:fmodel:lumped}
\partial_t \hv_{A_{00}} & = -\ha_1 \hv_{A_{00}} \hv_B + \ha_2 \hv_{A_{01}} \nonumber \\
\partial_t \hv_{A_{10}} & = \ha_1 \hv_{A_{00}} \hv_B - \ha_2 \hv_{A_{01}}
- \ha_3 \hv_{A_{01}} \hv_B + \ha_4 \hv_{A_{11}} \\
\partial_t \hv_{A_{11}} & = \ha_3 \hv_{A_{01}} \hv_B - \ha_4 \hv_{A_{11}} \nonumber \\
\partial_t \hv_{B} & = -\ha_1 \hv_{A_{00}} \hv_B + \ha_2 \hv_{A_{01}}
-\ha_1 \hv_{A_{00}} \hv_B + \ha_2 \hv_{A_{01}} , \nonumber
\end{align}
where $\ha \in [\hat{\lb};\hat{\ub}]$. Then, for any $\varepsilon, \delta > 0$, Theorem~\ref{thm:ucrn:lump:fluid} essentially implies that
\begin{itemize}
    \item for any $\alpha$, there is an $\ha$ such that, with a probability of $1 - \delta$ or higher, it holds that $|v^\alpha_{A_{01}} + v^\alpha_{A_{10}} - \hv^{\ha}_{A_{01}}| < \varepsilon$ and $|v^\alpha_S - \hv^{\ha}_S| < \varepsilon$ for all $S \notin \{A_{01},A_{10}\}$;
    \item for any $\ha$, there is an $\alpha$ such that, with a probability of $1 - \delta$ or higher, it holds that $|v^\alpha_{A_{01}} + v^\alpha_{A_{10}} - \hv^{\ha}_{A_{01}}| < \varepsilon$ and $|v^\alpha_S - \hv^{\ha}_S| < \varepsilon$ for all $S \notin \{A_{01},A_{10}\}$.
\end{itemize}
\end{example}

\begin{remark}
In contrast to Theorem~\ref{thm:ucrn:lump:2}, Theorem~\ref{thm:ucrn:lump:fluid} does not require the CCRN or its lumping to be regular. Rather, it requires that the reachable set $\frR$ of the CCRN does not exhibit an explosion on $[0;T]$, a property that can be often established for CRNs via conservation laws~\cite{10.1371/journal.pcbi.1004012,doi:10.1073/pnas.0308265100,DBLP:journals/bioinformatics/CardelliPTTVW21}. 
\end{remark}


Since CCRN species equivalence essentially ensures that the trajectories of the fluid models of the original and lumped CCRN coincide, it is natural to extend Theorem~\ref{thm:ucrn:lump:fluid} to value functions.

\begin{theorem}[Value preservation]\label{thm:ucrn:lump:fluid:costs}
Additionally to the assumptions made in Theorem~\ref{thm:ucrn:lump:fluid}, introduce
\begin{itemize}
    \item the differentiable running cost $L : \RE^\calS \times \RE_{\geq0} \to \RE_{\geq0}$ and final cost $K : \RE^\calS \to \RE_{\geq0}$ and;
    \item the functional $J_\alpha(v[0]) \! = \! \int_0^T \! L(t,v^\alpha(t))dt \! + \! K(v^\alpha(T))$, where $\partial_t v^\alpha = f(v^\alpha,\alpha)$, $v(0) = v[0]$ and $\alpha \in [\lb ;\ub]$;
    \item assume that $\partial_{v_A} L = \partial_{v_B} L$ and $\partial_{v_A} K = \partial_{v_B} K$ for all $H \in \calH$ and $A,B \in H$.
\end{itemize}
With this, define for any $\hv \in \RE_{\geq0}^{\hS}$ the lumped costs as $\hL(t,\hv) = L(t,v)$ and $\hK(t,\hv) = K(t,v)$, where $v \in \RE_{\geq0}^\calS$ is arbitrary such that $\sum_{A \in H} v_A = \hv_{A_H}$ for all $H \in \calH$. Then, for any initial condition $v[0] \in I$, almost surely it holds that
\[
\inf_\alpha J_\alpha(v[0]) = \inf_{\ha} \hJ_{\ha}(\hv[0]) ,
\]
provided that $\sum_{A \in H} v_A[0] = \hv_{A_H}[0]$ for all $H \in \calH$. A similar statement holds true for $\sup$.
\end{theorem}

\begin{proof}
The assumption on the running and final cost ensure that $L(t,v) = \hL(t,\hv)$ and $\hK(t,\hv) = K(t,v)$ for all $\hv \in \RE_{\geq0}^{\hS}$ and all $v \in \RE_{\geq0}^\calS$ satisfying $\sum_{A \in H} v_A = \hv_{A_H}$ for all $H \in \calH$. Since  $\frR([0;T]) \subseteq B(c)$ and $L$, $K$ are Lipschitz continuous on $B(c)$ as differentiable functions, Theorem~\ref{thm:ucrn:lump:fluid} ensures that for any initial condition $v[0] \in I$ and $\varepsilon, \delta > 0$, we have that
\[
\mathbb{P}(E(\varepsilon)) = \mathbb{P}\big\{ \big| \inf_\alpha J_\alpha(v[0]) - \inf_{\ha} \hJ_{\ha}(\hv[0]) \big| > \varepsilon \big\} < \delta ,
\]
provided that $\sum_{A \in H} v_A[0] = \hv_{A_H}[0]$ for all $H \in \calH$. Since this implies that $\mathbb{P}(E(\tfrac{1}{n})) < \tfrac{1}{n^2}$ for all $n \geq 1$, the Borel-Cantelli lemma ensures that
\[
\mathbb{P}\big\{ \big| \inf_\alpha J_\alpha(v[0]) - \inf_{\ha} \hJ_{\ha}(\hv[0]) \big| > 0 \big\} = 0 ,
\]
thus yielding the claim.
\end{proof}

\subsection{Reconstruction of Optimal Controls}

Thanks to Theorem~\ref{thm:ucrn:lump:fluid:costs}, we know that the original and the lumped system coincide on the optimal costs. The next result describes how an optimal control of the original system can be reconstructed from an optimal control of the lumped system.

\begin{theorem}[Control Reconstruction]\label{thm:construct}
Let us fix a CCRN $(\calS,\calR_{[\lb;\ub]})$, a CCRN lumpability $\calH$ and let $(\hS,\hR_{[\hat{\lb};\hat{\ub}]})$ be the lumped CCRN. Further, let $c > 0$ and $T > 0$ be such that $\frR(t) \subseteq B(c)$ for any $t \in [0;T]$. Then, for any $\hat{a} \in [\hat{\lb};\hat{\ub}]$, $v \in B(c)$ and $\hv$ such that $\hv_{A_H} = \sum_{A \in H} v_A$ for all $H \in \calH$, it holds that
\[
\min_{a \in [\lb ; \ub]} \max_{H \in \calH} | \sum_{A \in H} f_A(v,a) - \hf_{A_H}(\hv,\hat{a}) | = 0
\]
Additionally, for any optimal solution $\partial_t \hv = \hf(\hv,\ha)$ of the lumped system, $\partial_t v^\ast(t) = f(v^\ast(t),a(v^\ast(t),t))$ is an optimal solution of the original system, where
\begin{align*}
a(v,t) & := \argmin_{a \in [\lb;\ub]} \sum_{H \in \calH} \Big\lVert \sum_{A \in H} f_A(v,a) - \hf_{A_H}(\hv(t),\ha(t)) \Big\rVert_2^2
\end{align*}
can be computed by means of convex quadratic programming in polynomial time.
\end{theorem}

\begin{proof}
We begin with the first statement, writing $v(0)$ and $\hv(0)$ for $v$ and $\hv$, respectively. Thanks to continuity, we can assume without loss of generality that $v(0)$ is from the interior of $B(c)$. Following the argumentation from Theorem~\ref{thm:ucrn:lump:fluid:costs} that invokes the Borel-Cantelli lemma, it suffices to prove that $\mathbb{P}(E(\eta)) < \delta$ for any $\eta, \delta > 0$, where event $E(\eta)$ is
\[
E(\eta) = \{ \omega \mid \min_{a \in [\lb ; \ub]} \max_{H \in \calH} | \sum_{A \in H} f_A(v,a) - \hf_{A_H}(\hv,\hat{a}) | > \eta \} .
\]
To this, end we set $\varepsilon = \eta^2$ and pick, using Theorem~\ref{thm:ucrn:lump:fluid} for $T = \eta$, $v(0)$ and $\ha(\cdot) = \hat{a}$, some $\alpha(\cdot) \in [\lb ; \ub]$ such that $\partial_t v^\ast = f(v^\ast,\alpha)$ and $\partial_t \hv = \hf(\hv,\ha)$ satisfy $\mathbb{P}(E'(\eta)) < \delta$, where $v^\ast(0) = v(0)$ and
\[
E'(\eta) = \{ \omega \mid \max_{H \in \calH} \sup_{0 \leq t \leq \eta} | \sum_{A \in H} v^\ast_A(t) - \hv_{A_H}(t) | > \eta \}
\]
(Note that by picking $\eta > 0$ sufficiently small, we can always ensure that $v$ remains in $B(c)$ because $v(0)$ is from its interior.) For event $E'(\eta)$, we note that any $a \in [\lb ; \ub]$ satisfies
\begin{multline*}
\max_{H \in \calH} \Big| \sum_{A \in H} f_A(v(0),a) - \frac{\hv_{A_H}(\eta) - \hv_{A_H}(0)}{\eta} \Big| \leq \mathcal{O}\big(\frac{\epsilon}{\eta}\big) \\
+ \max_{H \in \calH} \Big| \sum_{A \in H} f_A(v(0),a) - \frac{1}{\eta} \Big( \sum_{A \in H} v^\ast_A(\eta) - \sum_{A \in H} v^\ast_A(0)\Big) \Big| \\
= \mathcal{O}(\eta) + \max_{H \in \calH} \Big| \sum_{A \in H} \Big(f_A(v(0),a) - f_A(v^\ast(t'_A),\alpha(t'_A)) \Big) \Big| \\
\leq \mathcal{O}(LC\eta) + \max_{H \in \calH} \Big| \sum_{A \in H} f_A(v(0),a) - \sum_{A \in H} f_A(v(0),\alpha(0)) \Big| ,
\end{multline*}
where $C$ and $L$ are, respectively, on $B(c) \times [\lb;\ub]$ an upper bound and a Lipschitz constant of each $f_A$, while each time point $t'_A \in [0 ; \eta]$ is picked via the mean value theorem. Overall, this implies that there exists a constant $K_1 \geq 0$, non dependent on $\eta$, such that
\[
\max_{H \in \calH} \Big| \sum_{A \in H} f_A(v(0),\alpha(0)) - \frac{\hv_{A_H}(\eta) - \hv_{A_H}(0)}{\eta} \Big| \leq K_1 \eta
\]
Moreover, applying Lagrange's form of Taylor's theorem to the function $t \mapsto \hv(t)$ ensures the existence of some $K_2 \geq 0$, non dependent on $\eta$, such that
\[
\max_{H \in \calH} \Big| \frac{\hv_{A_H}(\eta) - \hv_{A_H}(0)}{\eta} - f_{A_H}(\hv(0),\hat{a}) \Big| \leq K_2 \eta^2
\]
Overall, the discussion implies
\[
\min_{a \in [\lb ; \ub]} \max_{H \in \calH} \Big| \sum_{A \in H} f_A(v(0),a) - f_{A_H}(\hv(0),\hat{a}) \Big| \leq K_3 \eta
\]
for some $K_3 \geq 0$ that does not depend on $\eta$. Taking $\eta \to 0$ yields then the first statement. We next sketch the proof of the second statement. To this end, we approximate $a(\cdot,\cdot)$ by a Lipschitz continuous function $\bar{a}(\cdot,\cdot)$ given by $\bar{a}(v,t) := a(v,t)$ if $v$ is a point of a grid with mesh size $\tau > 0$, that is
\[
v \in G_\tau = B(c) \cap \{ (i \cdot \tau, j \cdot \tau ) \mid (i,j) \in \mathbb{Z} \times \mathbb{Z} \} ;
\]
for $v \notin G_\tau$, instead, we define $\bar{a}$ via an interpolation which ensures Lipschitzianity of $\bar{a}$, e.g., as a weighted sum of values at grid points most closest to $v$. Thanks to the definition of $a$ and the fact that $B(c)$ is a compactum, it can be shown that there exists a $K_4 \geq 0$, non dependent on $\tau$, such that
\begin{align*}
\sup_{\substack{v \in B(c)\\ t \in [0;T]}} \sum_{H \in \calH} \Big| \sum_{A \in H} f_A(v,a(v,t)) - \sum_{A \in H} f_A(v,\bar{a}(v,t)) \Big| \leq K_4 \tau
\end{align*}
Since $\bar{a}(\cdot,\cdot)$ is Lipschitz continuous on $B(c) \times [0 ; T]$, there exists a unique solution of $\partial_t v^\star(t) = f(v^\star(t),\bar{a}(v^\star(t),t))$. Moreover, the theory of differential equations (e.g., Gronwall's inequality) ensures that there is a constant $K_5 \geq 0$, non dependent on $\tau$, such that $\lVert \sum_{A \in H} v^\star_A(t) - \hv_{A_H}(t) \rVert_2 \leq K_5 \tau$ for all $H \in \calH$ and $0 \leq t \leq T$. This completes the proof.
\end{proof}

\section{Evaluation}\label{sec:eval}

We apply our framework to two families of models, one from epidemiology (and networks), and one from biology. The former class is used as an example for control and reachability, while the latter for reachability only. Algorithm~\ref{algorithm_part} has been implemented in ERODE~\cite{tacas2017} which supports~\cite{DBLP:journals/bioinformatics/CardelliPTTVW21}. The experiments were run on a 3.22 GHz machine assigning 6 GB of RAM to ERODE. In all cases, two iterations of Algorithm~\ref{algorithm_part} were sufficient.

\begin{table*}[th!]
	\centering
	\scalebox{0.9}{
		\begin{tabular}{crrrrrrrrrr}
			\toprule
			\multicolumn{11}{c}{\emph{\textbf{Spatial SIR with vaccination} for $\calG = \{ \{S_1\}, $ $\{I_1\},$ $\{R_1\},$ $\{ \text{remaining variables}\}\}$. Reductions have $7$ state variables.}}\\
			\midrule
			$n$ & 5000 & 10000 & 15000 & 20000 & 25000 & 30000 & 35000 & 40000 & 45000 & 50000 \\
			State variables & 20000 & 40000 & 60000 & 80000 & 100000 & 120000 & 140000 & 160000 & 180000 & 200000 \\
			Lumping time (ms) & 82 & 301 & 421 & 618 & 622 & 754 & 1045 & 1192 & 1276 & 1824 \\
			\bottomrule
		\end{tabular}
	}
	\caption{Running times of Algorithm~\ref{algorithm_part} for SIR on star networks.}\label{table:runtimesSIRStar}
\end{table*}

\subsection{SIR models over Networks}\label{eval:star}


\paragraph*{Scalability analysis} Disease spread over networks is often modeled by variants of the susceptible-infected-recovered (SIR) model~\cite{pastor2015epidemic} over graphs~\cite{DBLP:journals/bioinformatics/CardelliPTTVW21}. Here, we study an SIR variant with vaccination~\cite{SIRwithVac} over a star topology with $n$ locations ($5000$ to $50000$ with step $5000$ in Table~\ref{table:runtimesSIRStar}). The respective reactions are
\begin{align*}
S_i  &  \act{[\lb ; \ub]} R_i + V_i, &
S_i + I_j  &  \act{b_{ij} \beta} I_i + I_j,  \\
I_i  & \act{\gamma} R_i, &
R_i  & \act{\eta} S_i ,
\end{align*}
where $i,j \in \{1,\ldots,n\}$. The first reaction models the vaccination, the second captures the infection across different locations, the third recovery, while the forth corresponds to the loose of immunity. Subscripts denote locations and $B = (b_{i,j})$ is the adjacency matrix of the graph representing the network topology, with $b_{ij} > 0$ denoting the presence of an 
edge between node $i$ and $j$. The auxiliary species $V_i$ keep track of the vaccinated. Parameters $\beta, \gamma, \eta$ were chosen as in~\cite{DBLP:journals/bioinformatics/CardelliPTTVW21}, while vaccination bounds were set to $\lb = 0$ and $\ub = 1$ for lack of better alternative. By identifying $i = 1$ as the center of the network, the aforementioned star topology can be realized by setting $b_{1,i} = b_{i,1} = 1$ for all
$i \in \{2,\ldots,n\}$, and $b_{k,l}=0$ for all other cases. This means that infections can occur only among the center node and the others, further preventing infections within the same location. 
The intuition is that each node represents an individual that can get infected by interacting with others accordingly to the network topology.
We applied our lumping algorithm starting from the initial partition $\calH = \{ \{S_1\}, $ $\{I_1\},$ $\{R_1\},$ $\{V_1,\ldots,V_n\}$, $\{ \text{\emph{remaining variables}}\}\}$. Its lumped CCRN is given by the same reactions, but for $n = 2$. The original star topology with $n$ locations is thus reduced to one with $2$ locations. As a possible cost, one can consider the cumulative population of infected and vaccinated over time, that is 
\begin{align*}
J = \omega_1 \int_0^T \sum_{i = 1}^n v_{I_i}(s) ds + \omega_2 \sum_{i = 1}^n v_{V_i}(T) ,
\end{align*}
where $\omega_1$ and $\omega_2$ are non-negative weights. Intuitively, the cost aims at minimize the spread of infection while using a minimal amount of vaccination. With this, Theorem~\ref{thm:ucrn:lump:fluid:costs} ensures that the optimal value of the original CCRN of size $4n$ can be obtained by optimizing the lumped one of size $7$.

Overall, Table~\ref{table:runtimesSIRStar} shows that, for the considered family of models, the runtime of our technique scales well with the model size, taking less than 2 seconds for a model with 150 thousand variables.

\paragraph*{Reduction power analysis} Here we perform an analysis akin to the one in the foregoing paragraph. Specifically, we study the SIR model with vaccination, but this time over real-world networks taken from the Netzschleuder repository~\cite{Netzschleuder}. The rationale is that, after having studied runtime performances of CCRN species equivalence, here we focus on the reduction power of our technique in realistic settings.
We consider all weighted networks from the repository with at most 52000 nodes. Part of the networks are directed, while the others are undirected. We implicitly transform the latter ones by replacing every undirected edge with two corresponding directed ones with same weight. Overall, we considered 1558 real networks from the repository. The largest considered network contains 51919 nodes, corresponding to a CCRN with 155757 variables and 330149 reactions on which our reduction algorithm took about 50 minutes on a standard laptop machine.

\begin{figure*}[t]
		\subfigure[The 877 reducible models sorted by reduction ratio.]{\includegraphics[width = 0.5\textwidth]{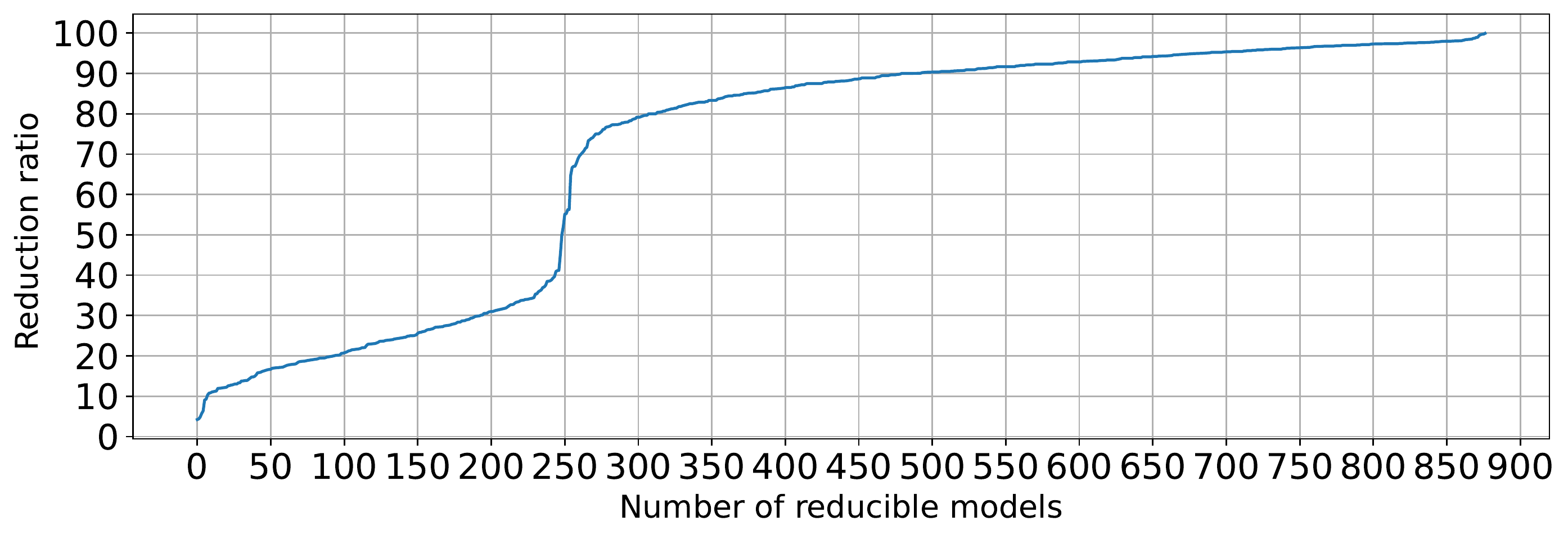}}
		\subfigure[All 1558 models grouped by reduction ratio. The gray numbers count the models in the corresponding range.\label{fig:weightedNetworksb}]{\includegraphics[width = 0.5\textwidth]{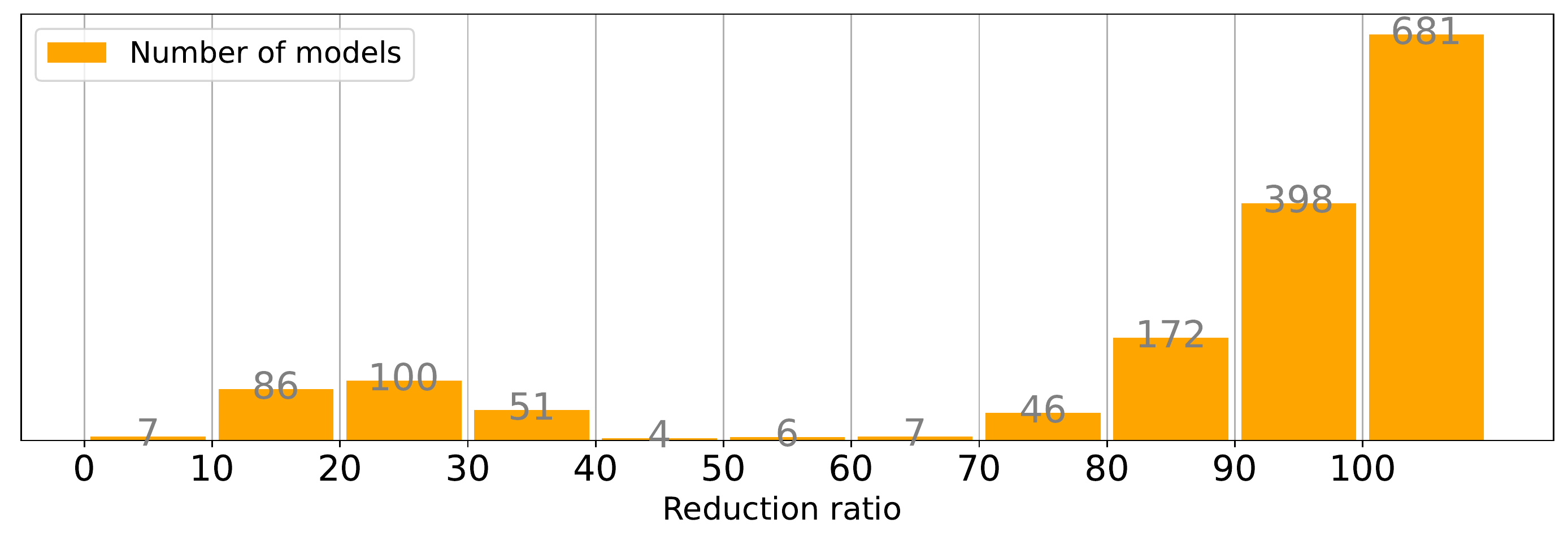}}
		\caption{CCRN lumping of SIR-vaccination model over weighted networks from~\cite{Netzschleuder}. Reduction ratios are given as number of reduced variables over original ones.\label{fig:weightedNetworks}}
	\end{figure*}
\begin{figure*}[t]
	\subfigure[Models without any uncertainty.\label{fig:weightedNetworks_UncertainWeightsa}]{\includegraphics[width = 0.5\textwidth]{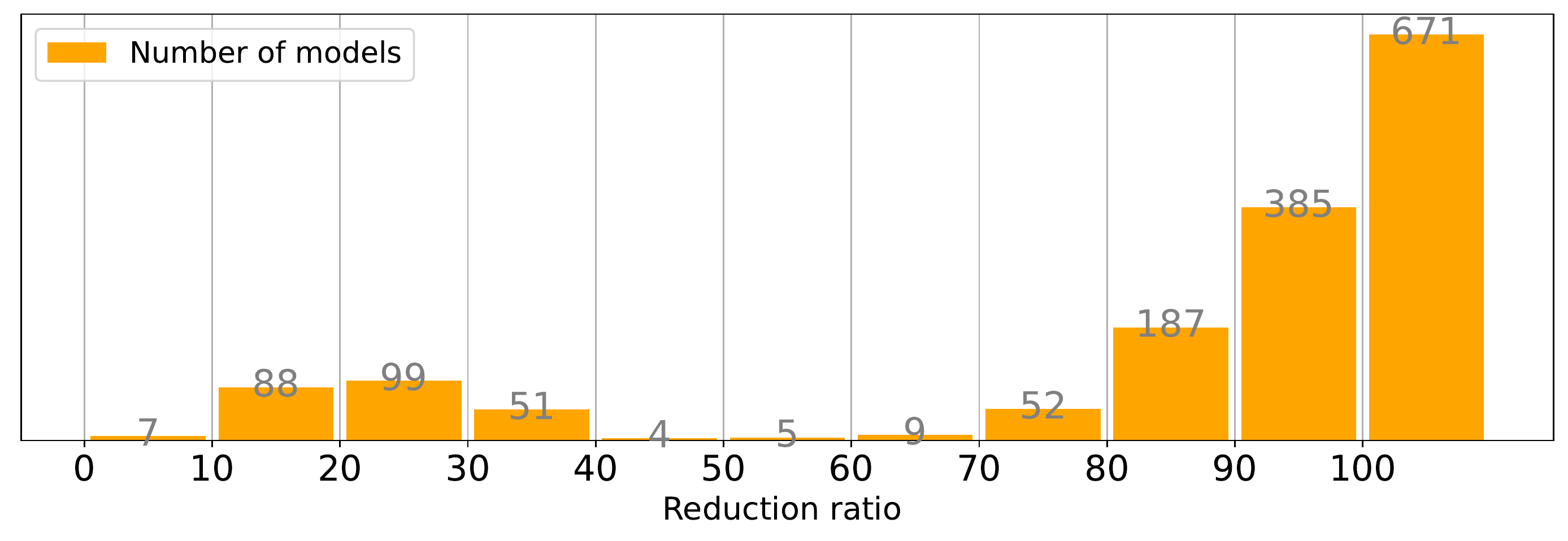}}
	\subfigure[Models with uncertainty on weights.\label{fig:weightedNetworks_UncertainWeightsb}]{\includegraphics[width = 0.5\textwidth]{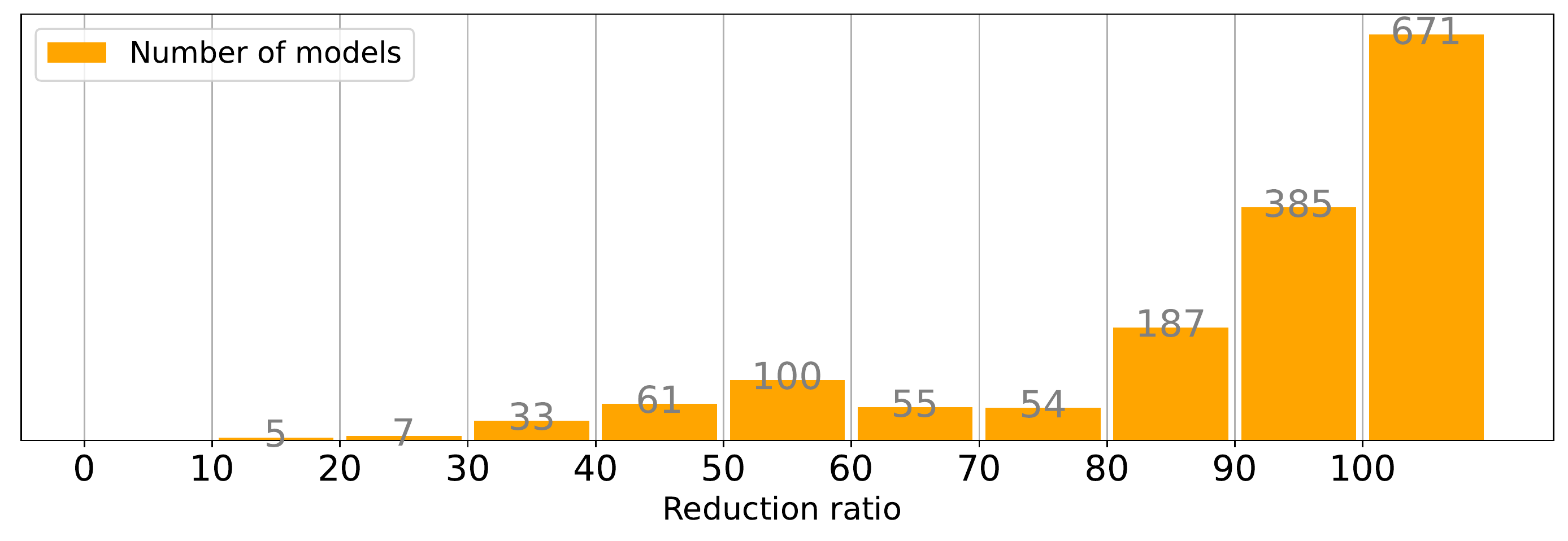}}
	\caption{
		CCRN lumping of SIR-vaccination model over weighted networks from~\cite{Netzschleuder}. Reduction ratios are given as number of reduced variables over original ones.
		All 1558 models grouped by reduction ratio. The gray numbers count the models in the corresponding range.
		\label{fig:weightedNetworks_UncertainWeights}}
\end{figure*}

All reactions apart the one modeling infections remain the same for the star topology, including the parameters and $\lb = 0$ and $\ub = 1$. As regards infections, for every edge from node $i$ to node $j$ with weight $w_{ij}$ we add a reaction
\[ S_j + I_i  \act{w_{ij}} I_j + I_i\]
In other words, in our experiment we interpret the presence of an edge from node $i$ to $j$ as the possibility for individual $i$ ($I_i$) to infect individual $j$ ($S_j$).
On these models, we applied our lumping algorithm starting from an initial partition with blocks that separate the types of variables across all nodes:
\[\calH = \{ \{ S_i \mid i \leq n\}, \{ I_i \mid  i \leq n\}, \{ R_i \mid  i \leq n\}, \{V_i \mid i \leq n \} \}\]
with $n$ the number of nodes in the considered network. The cost can be taken as in the case of the star topology. More generally, for any reported lumping $\calH$, any costs $L, K$ satisfying the assumptions of Theorem~\ref{thm:ucrn:lump:fluid:costs} are applicable, e.g., costs that try to minimize the cumulative infection in a specific block of $\calH$.

The results are summarized in Figure~\ref{fig:weightedNetworks}.
We define the reduction ratio of a model as the number of reduced variables over that of original ones (the auxiliary species $V_i$ have been dropped for providing a cleaner picture). Overall, 877 models could be reduced (i.e., have a reduction ratio smaller than 1), while 681 were not reduced (reduction ratio = 1).
Figure~\ref{fig:weightedNetworks} (a) focuses on the 877 models that admitted reduction, sorted by reduction ratio. We can see that about 250 models could be reduced to less than half the original number of variables.
This is visualized better in Figure~\ref{fig:weightedNetworks} (b). Here we count how many models have a reduction ratio within ten intervals from [0.0;0.1] (the bar from 0 to 10), to [0.9;1.0] (the bar from 90 to 100). The right-most bar refers to the 681 models that did not admit any reduction.

Overall, more than 56\% of the models admitted reduction. Among these, about 28\% admitted substantial reductions obtaining a reduction ratio smaller than 0.4. 

\begin{table*}[th!]
	\centering
	\scalebox{0.9}{
		\begin{tabular}{crrrrrrrrrr}
			\toprule
			\multicolumn{11}{c}{\emph{\textbf{Protein-interaction networks} for $\calG = \{ \calS \}$. Reductions  have $n+1$ state variables.}}\\
			\midrule
			$n$ & 9 & 10 & 11 & 12 & 13 & 14 & 15 & 16 & 17 & 18 \\
			State variables & 513 & 1025 & 2049 & 4097 & 8193 & 16385 & 32769 & 65537 & 131073 & 262145 \\
			Lumping time (ms) & 63 & 81 & 88 & 96 & 253 & 430 & 841 & 2157 & 5024 & 10582 \\
			\bottomrule
		\end{tabular}
	}
	\caption{Running times of Algorithm~\ref{algorithm_part} on protein-interaction networks.}\label{table:runtimesBinding}
\end{table*}

\paragraph*{Impact of uncertain weights on reduction power} 
In this experiment, rather than focusing on the control problem of vaccination, we study the impact of weights' uncertainty on the reduction power of our technique. To this end, we perform a new analysis of the SIR vaccination model over networks from~\cite{Netzschleuder} by fixing the vaccination rate (to 1), while assuming that there is uncertainty in the weights of the 1558 networks considered (we use an arbitrary interval of $0.05$ centered at weights's values, to ensure that intervals remain positive).
The results are summarized in Figure~\ref{fig:weightedNetworks_UncertainWeights}.
Similarly to Figure~\ref{fig:weightedNetworksb}, we group models by reduction ratio. In particular, Figure~\ref{fig:weightedNetworks_UncertainWeightsa} considers models without uncertainty of the weights, while Figure~\ref{fig:weightedNetworks_UncertainWeightsb} with uncertainty on the weights.
We can see that the absolute number of reducible models is not affected (in both cases, 671 models could not be reduced at all). Likewise, mild reductions with reduction ratios  from 0.7 to 1.0 are not affected either.
Considering the cases with lower reduction ratio, we can clearly see a similar pattern in the two figures, shifted to the right in the case of uncertain weights: reduction ratios from 0.1 to 0.4 appear to get shifted from 0.3 to 0.7.

\subsection{Protein-interaction Networks}
Models of signaling pathways exhibit often a rapid growth in the number of species and reactions because of distinct molecule configurations~\cite{FEBS:FEBS7027,citeulike:8493139}. 
A possible instance of this is an extension of Example~\ref{ex:model} to $n$ binding sites ($9\leq n \leq 18$ in Table~\ref{table:runtimesBinding}), yielding species $\calS = \big\{ \{B\} , \{ A_b \mid b \in \{0,1\}^n \} \big\} $ and reactions
\begin{align*}
	A_b + B & \act{[\lb_a ; \ub_a]} A_{b + e_i} , & b_i & = 0 \\
	A_b & \act{[\lb_d ; \ub_d]} A_{b - e_i} + B , & b_i & = 1 ,
\end{align*}
where $e_i$ denotes the $i$-th unit vector, and the subscripts $a$ and $d$ denote association and disassociation with parameter bounds $[9.95;10.05]$ and $[0.05;0.15]$, respectively. The bounds are in accordance with the exact values $10$ and $0.1$ from~\cite{citeulike:8493139}.
Similarly to Example~\ref{ex:model} with two binding sites, it can be shown that $\calH = \big\{ \{ A_b \mid |b| = 0 \}, $ $ \ldots, $ $\{ A_b \mid |b| = n \}, $ $\{B\} \big\}$ is a CCRN species equivalence. The lumped CCRN can be then described by $\hS = \{ B , A_0, \ldots, A_n \}$ and the reactions
\begin{align*}
	A_i + B & \act{[\lb_a ; \ub_a]} A_{i+1} , & 0 \leq i & < n , \\
	A_i & \act{[\lb_d ; \ub_d]} A_{i - 1} + B , & 0 < i & \leq n .
\end{align*}
That is, similarly to Example~\ref{ex:model}, the CCRN species equivalences keep track of the number of occupied binding sites rather than the actual configuration of each binding site. The largest considered model has about 250000 variables, requiring about 10 seconds. The running times are summarized in Table~\ref{table:runtimesBinding}.

Overall, Theorem~\ref{thm:ucrn:lump:fluid:costs} ensures that the reachable set of the original CCRN of size $2^n + 1$ coincides with that of the lumped CCRN of size $n + 2$ on the blocks of $\calH$. The lumped CCRN can be over-approximated by known techniques like~\cite{DBLP:journals/automatica/Lygeros04,DBLP:conf/rtss/ChenAS12}.

\section{Conclusion}\label{sec:conc}

We introduced a model reduction technique for controlled chemical reaction networks (CCRNs) whose kinetic reaction parameters are subject to control or distrubance. The smallest (lumped) CCRN can be computed in polynomial time and is shown to preserve the optimal costs of the original CCRN. The applicability has been demonstrated by reducing the reachability and control problems of protein-interaction networks and vaccination models over networks with hundreds of thousands of variables. In the latter case, the runtime scalability has been demonstrated on synthetic networks with star topology, while the aggregation power in practice has been demonstrated by considering real-world weighted networks.

The proposed framework is holistic in that it can be used as a precomputation step before any optimization approach. In case the reduced model is sufficiently small, global optimization techniques such as the Hamilton-Jacobi-Bellman equations~\cite{DBLP:journals/automatica/Lygeros04,DBLP:conf/cdc/ChenT15} or reachability analysis tools such as~\cite{DBLP:conf/rtss/ChenAS12,DBLP:conf/cav/BogomolovFGLPW12,Althoff2015a} can be invoked. If the reduced model is still too large for global optimization techniques, local optimization approaches such as the functional gradient descent, also known as Pontryagin's maximum principle~\cite{Liberzon}, can be invoked. While the principle has gained recently momentum in AI by training so-called neural ordinary differential equations~\cite{DBLP:conf/nips/ChenRBD18}, its computational complexity is at least quadratic in the size of the model, thus justifying the need for optimality-preserving model reduction techniques. Likewise, heuristic approaches involving sampling and simulation, as commonly used in systems biology~\cite{10.1093/bioinformatics/btp619}, can profit from optimality-preserving reductions as well.

\section{Proof of Proposition~\ref{prop:limit}}\label{sec:postponed}

\begin{proof}
We denote a reaction $r = (\rho \act{[\lb_r;\ub_r]} \pi ) \in \calR[\lb;\ub]$ simply by $\rho \act{} \pi$ since the range of its reaction rate is clear from the context. For a given $N\geq1$, fix $q \in q^N_{[\lb;\ub]}$. For every $\sigma, \theta \in \frac{1}{N} \NA^\calS$ with $\theta \neq \sigma$, the $(\sigma,\theta)$ entry of $q$ is $q(\sigma,\theta) \in g(\sigma) \cdot [q_{\lb^N}(N\sigma,N\theta) ; q_{\ub^N}(N\sigma,N\theta) ]$, so it has the form that we now describe. For every reaction $r = (\rho \to \pi) \in \calR$ such that $\theta = \sigma + \frac{1}{N}(\pi - \rho)$ there is $\alpha_r = \alpha_r(N,\sigma,\theta,t) \in [\lb_r;\ub_r]$ with $t \in [0;T]$ such that
\begin{equation}\label{description:q}
q(\sigma,\theta) = g(\sigma) \cdot
\sum_{ \substack{ r = (\rho \act{} \pi ) \in \calR \\ \theta = \sigma + \frac{1}{N}(\pi - \rho) } } \frac{\alpha_r(N,\sigma, \theta,t)}{ N^{|\rho|-1} } \binom{N \sigma}{\rho} 
\end{equation}
In particular, for a given $\sigma \in \frac{1}{N} \NA^\calS$, one has $q(\sigma,\theta) \neq 0$ only for finitely many $\theta \in \frac{1}{N} \NA^\calS$. We begin the proof by checking that our scaled UCTMCs satisfy~\cite[Definition 4 (i)-(iii)]{dsn16BortolussiGast}, in order to then apply~\cite[Theorem 1]{dsn16BortolussiGast}.

(i) We show that for every $N$ we have
\[
s :=\sup_{\sigma \in \frac{1}{N} \NA^\calS} \sup_{ q \in q^N_{[\lb;\ub]} } | q(\sigma,\sigma) | < \infty.
\]
As $g(\sigma) = 0$ for every $\sigma \in \frac{1}{N} \NA^\calS$ with $|\sigma| \geq 2c$, we have
\begin{align*}
s &= \sup_{ \sigma \in \frac{1}{N} \NA^\calS } \sup_{ q \in q^N_{[\lb;\ub]} } \sum_{\sigma \neq \theta \in \frac{1}{N}\NA^\calS} q (\sigma,\theta) \\
&= \sup_{ \substack{ \sigma \in \frac{1}{N} \NA^\calS \\ |\sigma|\leq 2c } } g(\sigma) \cdot \sum_{\sigma \neq \theta \in \frac{1}{N}\NA^\calS}
q_{\ub^N}(N\sigma,N\theta).
\end{align*}
The number of elements $\sigma \in \frac{1}{N} \NA^\calS$ with $|\sigma|\leq 2c$ is finite, and for each of them the term
$\sum_{\sigma \neq \theta \in \frac{1}{N}\NA^\calS} q_{\ub^N}(N\sigma,N\theta)$
is a finite sum, so $s$ is finite.

(ii) For $\varepsilon \geq 0$, let
\[
\Phi_{ \varepsilon }(N)= 
\sup_{\sigma \in \frac{1}{N} \NA^\calS} \sup_{ q \in q^N_{[\lb;\ub]} } \sum_{\theta \in \frac{1}{N}\NA^\calS} q(\sigma,\theta) |\theta-\sigma|^{1+\varepsilon}.
\]
Here we prove that $\lim_{N \to \infty} \Phi_{ \varepsilon }(N) = 0$ for every $\varepsilon > 0$, and to this end it is sufficient to show that
\begin{equation}\label{statement:epsilon}
\text{$\Phi_{ \varepsilon }(N)$ is $O( N^{-\varepsilon} )$ as $N\to \infty$, for every $\varepsilon \geq 0$.}
\end{equation}
Fix $\varepsilon \geq 0$. As $g(\sigma) \leq 1$ for every $\sigma \in \frac{1}{N} \NA^\calS$, and $g(\sigma) = 0$ when $|\sigma| \geq 2c$, using \eqref{description:q} we obtain
\begin{align*}
& \hspace{-5pt}
 \Phi_{ \varepsilon }(N) \leq
\sup_{ \substack{ \sigma \in \frac{1}{N} \NA^\calS \\ |\sigma| \leq 2c , \\ \alpha \in [\lb;\ub] } } 
\sum_{ \substack{ r = (\rho \to \pi ) \in \calR \\ \sigma + \frac{\pi - \rho}{N} \in \frac{1}{N} \NA^\calS } } \frac{\alpha_r}{ N^{|\rho|-1} } \cdot \binom{N \sigma}{\rho} \cdot | \frac{\pi - \rho}{N} |^{1+\varepsilon}\\
&= \sup_{ \substack{ \sigma \in \frac{1}{N} \NA^\calS \\ |\sigma| \leq 2c } } \sum_{ \substack{ r = (\rho \to \pi ) \in \calR \\ \sigma + \frac{\pi - \rho}{N} \in \frac{1}{N} \NA^\calS } } \ub_r |\pi - \rho|^{1+\varepsilon} \cdot \prod_{A \in \calS} \frac{\binom{N \sigma(A)}{\rho(A)}}{ N^{\rho(A)} } \cdot N^{-\varepsilon}.
\end{align*}
It can be shown that for every $\rho \in \rho(\calR)$ and $\sigma \in \frac{1}{N} \NA^\calS$ with $|\sigma| \leq 2c$ one has
\begin{equation}\label{Kurtz:limit}
\frac{\binom{N \sigma(A)}{\rho(A)}}{ N^{\rho(A)} }  = \frac{ \sigma(A)^{ \rho(A) } }{ \rho(A)! } + O( N^{-1} ) \leq \frac{ (2c)^{ \rho(A) } }{ \rho(A)! } + O( N^{-1} ).
\end{equation}
Then one can find a constant $K\geq 0$ depending only on the set $\calR$ of reactions (and on $c$) such that for every reaction $(\rho \to \pi ) \in \calR$ 
one has
\[
|\pi - \rho|^{1+\varepsilon} \prod_{A \in \calS} \frac{1}{ N^{\rho(A)} } \binom{N \sigma(A)}{\rho(A)} \leq K + O( N^{-1} )
\]
for every $N$ that is big enough. Then for such $N$ we obtain
\[
\Phi_{ \varepsilon }(N) \leq \sum_{ r = (\rho \to \pi ) \in \calR } \ub_r \big( K + O( N^{-1} ) \big) N^{-\varepsilon}.
\]
As the right-hand side above is $O( N^{-\varepsilon}) + O( N^{-1-\varepsilon} ) = O( N^{-\varepsilon})$ for $N \to \infty$, we deduce \eqref{statement:epsilon}.

(iii) We have to check that
$\limsup_{N \to \infty} \Phi_{ 0 }(N) < \infty$, which readily follows from \eqref{statement:epsilon}. This finishes the proof ~\cite[Definition 4 (i)-(iii)]{dsn16BortolussiGast}.

To apply~\cite[Theorem 1]{dsn16BortolussiGast}, we also need to show that the drifts $f^N$ of the CTMCs $(\frac{1}{N} \NA^\calS, q^N)$, where $q^N \in q^N_{[\lb;\ub]}$, describe for $N \to \infty$ the (upper semicontinuous) differential inclusion
\[
F(v) = \bigcup_{\alpha \in [\lb;\ub]} g(v) \cdot f(v,\alpha)
\]
For this, fix $q \in q^N_{[\lb;\ub]}$ and $\sigma \in \frac{1}{N}\NA^\calS$. Then \eqref{description:q} gives
\begin{align*}
& f^N(\sigma, q) = \sum_{\theta \in \frac{1}{N}\NA^\calS} q (\sigma,\theta) (\theta-\sigma)\\
&= g(\sigma) \cdot \sum_{ 
r = (\rho \to \pi ) \in \calR }
\frac{1}{N}(\pi - \rho) \cdot \frac{\alpha_r}{ N^{|\rho|-1} } \cdot \prod_{A \in \calS} \binom{N \sigma(A)}{\rho(A)} \\
&= g(\sigma) \cdot \sum_{r = (\rho \to \pi ) \in \calR} (\pi - \rho) \cdot \alpha_r \cdot \prod_{A \in \calS} \frac{\binom{N \sigma(A)}{\rho(A)}}{ N^{\rho(A)} }.
\end{align*}
Then the drift of the UCTMC $X_N = (\frac{1}{N} \NA^\calS,q^N_{[\lb;\ub]})$ at $\sigma$ is
\begin{align*}
& \bigcup_{q \in q^N_{[\lb;\ub]}} f^N(\sigma, q)
= \bigcup_{\alpha \in [\lb;\ub]} f^N(\sigma, q) \\
&= g(\sigma) \cdot \bigcup_{\alpha \in [\lb;\ub]} \sum_{r = (\rho \to \pi ) \in \calR} (\pi - \rho) \cdot \alpha_r \cdot \prod_{A \in \calS} \frac{\binom{N \sigma(A)}{\rho(A)}}{ N^{\rho(A)} }.
\end{align*}
Now fix $v \in \RE_{\geq0}^\calS$ and let $\sigma^N \in \frac{1}{N}\NA^\calS$ be such that $\lim_{N \to \infty} \sigma^N = v$. As the function $g$ is continuous, we have $\lim_{N \to \infty} g(\sigma^N) = g(v)$. Similarly to \eqref{Kurtz:limit}, for any $\rho \in \rho(\calR)$
\[
\lim_{N \to \infty} \frac{ \binom{N \sigma^N(A)}{\rho(A)} }{ N^{\rho(A)} } = \frac{ v_A^{ \rho(A) } }{ \rho(A)! },
\]
so the limit drift as $N \to \infty$ is
\begin{align*}
& \lim_{N \to \infty } \bigcup_{q \in q^N_{[\lb;\ub]}} f^N(\sigma^N, q) = \\
& \hspace{-10pt} = \lim_{N \to \infty } g(\sigma^N) \cdot \bigcup_{\alpha \in [\lb;\ub]} \sum_{r = (\rho \to \pi ) \in \calR} (\pi - \rho) \cdot \alpha_r \cdot \prod_{A \in \calS} \frac{ \binom{N \sigma^N(A)}{\rho(A)} }{ N^{\rho(A)} } \\
& = g(v) \cdot \bigcup_{\alpha \in [\lb;\ub]} \sum_{r = (\rho \to \pi ) \in \calR} (\pi - \rho) \cdot \alpha_r \cdot \prod_{A \in \calS} \frac{ v_A^{ \rho(A) } }{ \rho(A)! } \\
& = \bigcup_{\alpha \in [\lb;\ub]} g(v) \cdot f(v,\alpha).
\end{align*}
The above discussion allows us to apply~\cite[Theorem 1]{dsn16BortolussiGast} which, in turn, yields 1), see also~\cite[Theorem 3.2]{doi:10.1137/110844192}.
\end{proof}

\bibliographystyle{IEEEtran}
\bibliography{cdc2022}




\vspace{-2.0cm}

\begin{IEEEbiography}[{\includegraphics[width=1in,height=1.5in,clip,keepaspectratio]{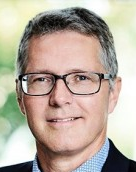}}]{Kim G. Larsen} is a Professor in Computer Science at Aalborg University and the director of the ICT-competence center CISS, Center for Embedded Software Systems. He is the holder of the VILLUM Investigator grant S4OS and led the ERC Advanced Grant LASSO. Moreover, he is the director of the Danish Innovation Network InfinIT, as well as the Innovation Fund Denmark research center DiCyPS.
\end{IEEEbiography}

\vspace{-2.0cm}

\begin{IEEEbiography}[{\includegraphics[width=1in,height=1.5in,clip,keepaspectratio]{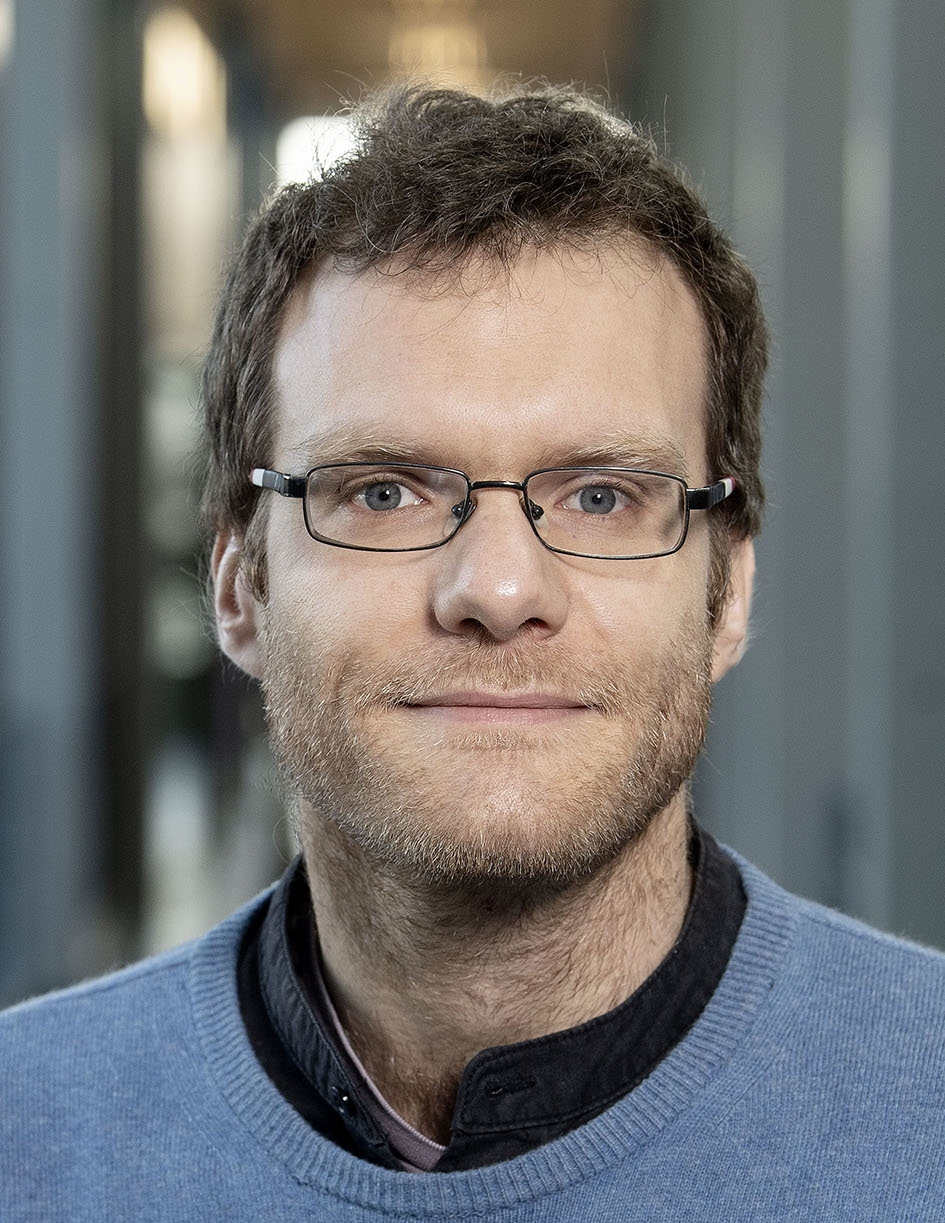}}]{Daniele Toller}
is a PostDoc Researcher at Aalborg University, Denmark. Prior to it, he was a PostDoc Researcher at the University of Udine, Italy, and at the University of Camerino, Italy. He has a Master Degree and a Ph.D. in Mathematics, received from the University of Udine.
\end{IEEEbiography}

\vspace{-2.0cm}

\begin{IEEEbiography}[{\includegraphics[width=1in,height=1.25in,clip,keepaspectratio]{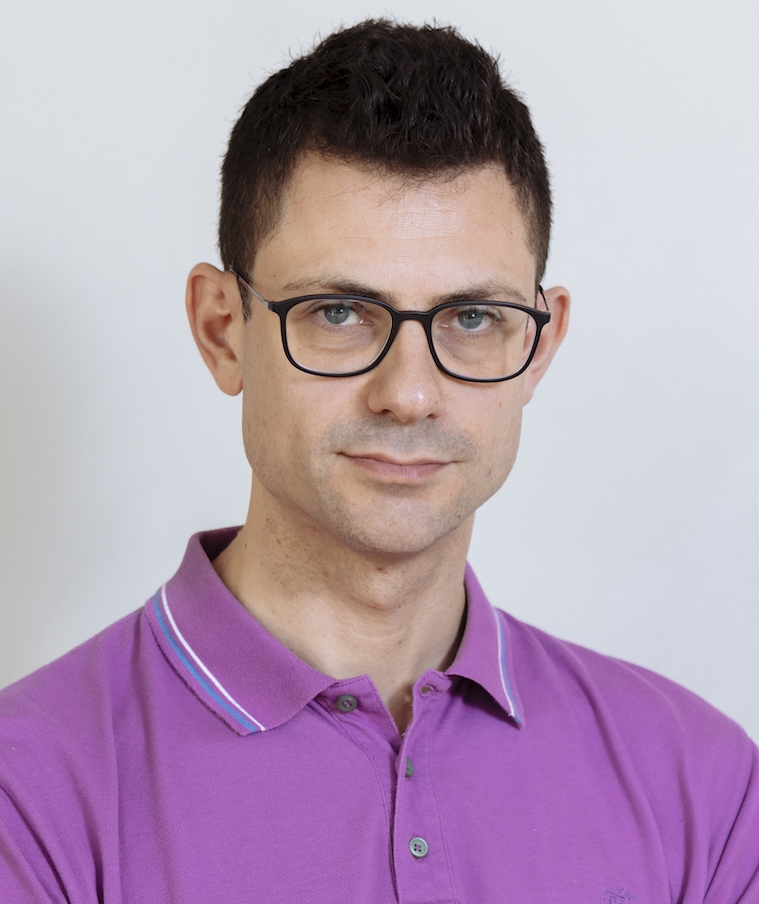}}]{Mirco Tribastone}
is a Professor at IMT Lucca, Italy. Prior to joining IMT Lucca he was Associate Professor at the University of Southampton, UK, and Assistant Professor at the Ludwig-Maximilians University of Munich, Germany. He received his Ph.D. in Computer Science from the University of Edinburgh, UK, in 2010. He graduated in Computer Engineering at the University of Catania, Italy.
\end{IEEEbiography}

\vspace{-2.0cm}

\begin{IEEEbiography}[{\includegraphics[width=1in,height=1.25in,clip,keepaspectratio]{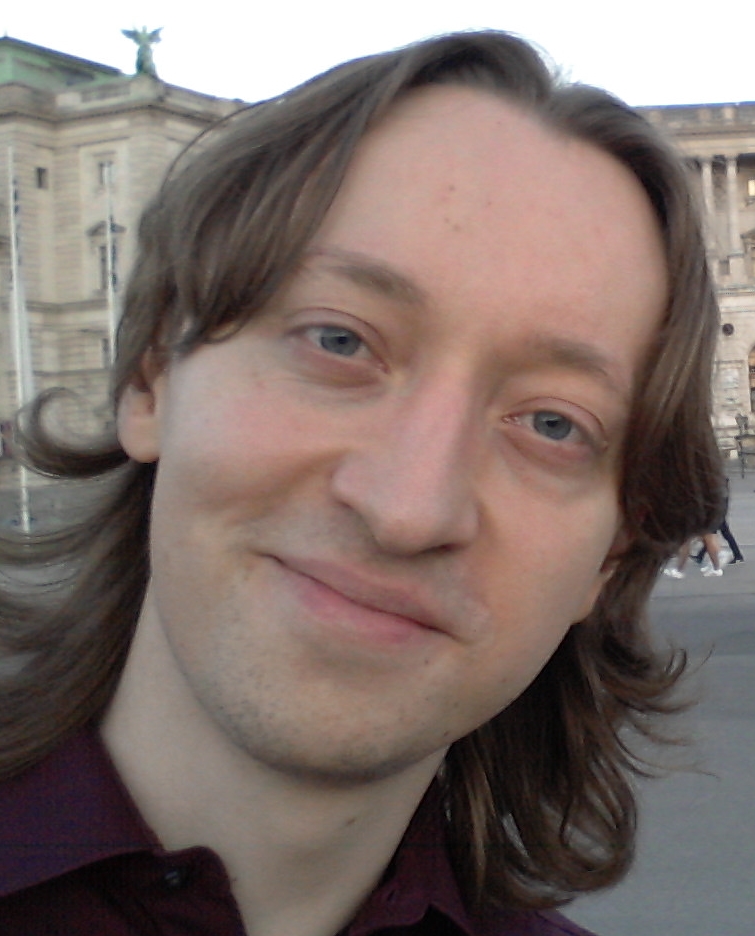}}]{Max Tschaikowski}
is a Poul Due Jensen Associate Professor at Aalborg University, Denmark. Prior to it, he was a Lise Meitner Fellow at TU Wien, Austria, an Assistant Professor at IMT Lucca, Italy, a Research Fellow at the University of Southampton, UK, and a Research Assistant at the Ludwig-Maximilians University in Munich, Germany. He was awarded a Diplom in mathematics (equivalent to a Master) and a Ph.D. in computer science by the LMU in 2010 and 2014, respectively.
\end{IEEEbiography}

\vspace{-2.0cm}

\begin{IEEEbiography}[{\includegraphics[width=1in,height=1.25in,clip,keepaspectratio]{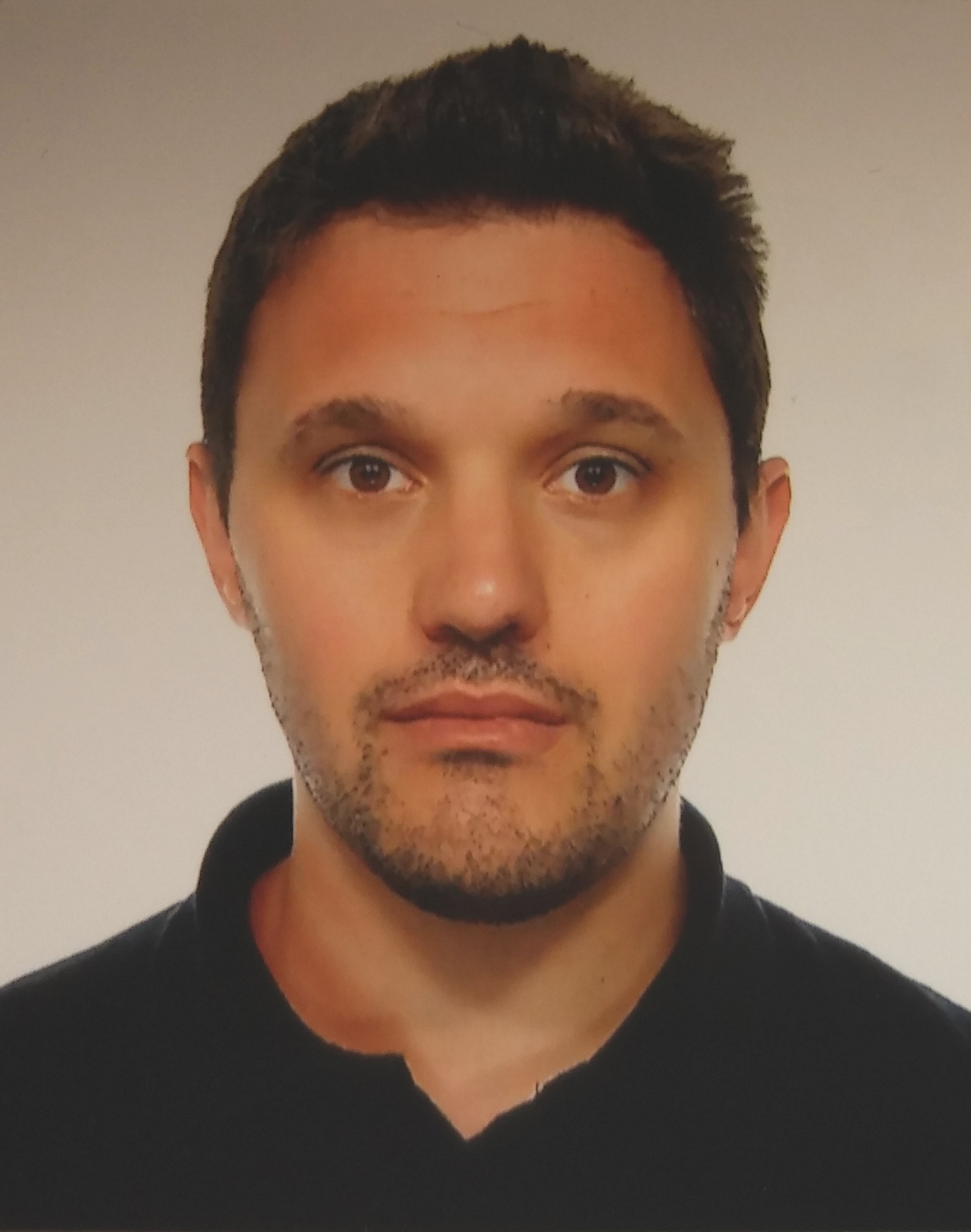}}]{Andrea Vandin}
is a tenure-track Assistant Professor at Sant'Anna School for Advanced Studies, Pisa, Italy, and an Adjunct Associate Professor at DTU Technical University of Denmark. Prior to it he was an Associate Professor at DTU and an Assistant Professor at IMT Lucca, Italy. In 2013-2015 he was a Senior Research Assistant at the University of Southampton, UK. He received his PhD in Computer Science and Engineering from IMT Lucca, Italy. He graduated in Computer Science at the University of Pisa, Italy.
\end{IEEEbiography}

\end{document}